\documentclass[aip]{revtex4-1}
\usepackage[version=3]{mhchem} 
\usepackage[caption=false]{subfig}
\usepackage{graphicx} 
\usepackage{dcolumn}

\usepackage{amssymb, amsmath}
\usepackage{braket}
\usepackage{bm}
\usepackage{upgreek}
\newcommand{\mr}{\mathrm}

\usepackage[authormarkuptext=name,authormarkup=none,final]{changes}
\definecolor{red}{RGB}{255, 0, 0}
\definecolor{purple}{RGB}{127, 0, 255}
\definecolor{blue}{RGB}{0, 0, 255}
\definecolor{green}{RGB}{0, 192, 0}
\definechangesauthor[name={Authors},color={red}]{GL1}
\definechangesauthor[name={Gianluca2}, color={purple}]{GL2}
\definechangesauthor[name={Benni}, color={blue}]{BN}

\makeatletter
\def\@email#1#2{%
 \endgroup
 \patchcmd{\titleblock@produce}
  {\frontmatter@RRAPformat}
  {\frontmatter@RRAPformat{\produce@RRAP{*#1\href{mailto:#2}{#2}}}\frontmatter@RRAPformat}
  {}{}
}%

\draft 

\begin{document}

\title{Decoherence and vibrational energy relaxation of the electronically excited PtPOP complex in solution} 

\author{Benedikt O. Birgisson}
\affiliation{Science Institute and Faculty of Physical Sciences, University of Iceland, Reykjavík, Iceland}
\author{Asmus Ougaard Dohn}
\affiliation{Department of Physics, Technical University of Denmark, 2800 Lyngby, Denmark}
\author{Hannes Jónsson} 
\affiliation{Science Institute and Faculty of Physical Sciences, University of Iceland, Reykjavík, Iceland}
\affiliation{Deptartment of Chemistry, Brown University, Providence, Rhode Island 02912, USA}
\author{Gianluca Levi}
\affiliation{Science Institute and Faculty of Physical Sciences, University of Iceland, Reykjavík, Iceland}
\email{giale@hi.is}

\begin{abstract}
Understanding the ultrafast vibrational relaxation following photoexcitation of molecules in a condensed phase is essential to predict the outcome and improve the efficiency of photoinduced molecular processes. Here, the vibrational decoherence and energy relaxation of a binuclear complex, [Pt$_2$(P$_2$O$_5$H$_2$)$_4$]$^{4-}$ (PtPOP), upon electronic excitation in liquid water and acetonitrile are investigated through direct adiabatic dynamics simulations. A quantum mechanics/molecular mechanics (QM/MM) scheme is used where the excited state of the complex is modelled with orbital-optimized density functional calculations while solvent molecules are described using potential energy functions. The decoherence time of the Pt-Pt vibration dominating the photoinduced dynamics is found to be $\sim$1.6 ps in both solvents. This is in excellent agreement with experimental measurements in water, where intersystem crossing is slow ($>10$ ps). Pathways for the flow of excess energy are identified by monitoring the power of the solvent on vibrational modes. The latter are obtained as generalized normal modes from the velocity covariances, and the power is computed using QM/MM embedding forces. Excess vibrational energy is found to be predominantly released through short-range repulsive and attractive interactions between the ligand atoms and surrounding solvent molecules, whereas solute-solvent interactions involving the Pt atoms are less important. Since photoexcitation deposits most of the excess energy into Pt-Pt vibrations, energy dissipation to the solvent is inefficient. This study reveals the mechanism behind the exceptionally long vibrational coherence of the photoexcited PtPOP complex in solution and underscores the importance of short-range interactions for accurate simulations of vibrational energy relaxation of solvated molecules.
\end{abstract}

\maketitle 

\section{Introduction}
The world's growing demand for economically viable and sustainable energy solutions has led to an increased interest in exploiting photoactive molecules for solar harvesting applications. When a molecule absorbs light, its vibrational degrees of freedom can be excited as a result of changes in the molecular potential energy surface upon electronic excitation. For a short pulse of light, a coherent superposition of vibrational states can be formed and coherent vibrations of the atoms can be observed following photoexcitation\cite{Zewail2000}. In a condensed phase, such as a solution, the coherent oscillations are suppressed by elastic and inelastic collisions with the environment, resulting in energy relaxation and randomization of the phase along the initially excited vibrational degrees of freedom. This process typically occurs on a time scale ranging from a few femtoseconds to a few picoseconds. Such transient photoinduced coherent vibrations in molecules are increasingly being detected in ultrafast pump-probe experiments, providing insights into the underlying potential energy surfaces of excited electronic states and the interactions with the environment\cite{Solling2018}. The role of coherent vibrations in energy conversion processes is also becoming more evident. For example, recent studies suggest that vibrational and electronic-vibrational (vibronic) coherences may enhance charge and energy transfer in photosynthetic systems\cite{Romero2017, Abramavicius2016, Chenu2015, Fuller2014}. Despite their importance and ubiquity in time-resolved studies, the mechanisms of decoherence and vibrational energy relaxation often remain unknown. Indeed, it is typically challenging, if possible at all, to disentangle pure dephasing effects and the pathways of intramolecular energy redistribution and dissipation to the environment from the observables of ultrafast experiments.

Bimetallic d$^8$-d$^8$ complexes, particularly Pt(II)-Pt(II) complexes, represent a class of photoactive molecules where electronic excitation induces long-lasting coherent atomic vibrations \cite{Yan2023, Chaaban2019, gray2017}. In Pt(II)-Pt(II) compounds, the lowest lying triplet and singlet excited states involve excitation from a metal-metal d$\sigma^*$ antibonding HOMO to a metal-metal p$\sigma$ bonding or ligand-centered LUMO, depending on the ligands. Depopulation of a metal-metal antibonding orbital upon photoexcitation leads to a shortening of the Pt-Pt bond, thereby launching coherent vibrations along the Pt-Pt coordinate\cite{Rafiq2023, Leshchev2023, Kim2022, Kim2021, Haldrup2019, monni2018, Veen2011}. The decay of these coherent oscillations and the timescale of intersystem crossing from the lowest singlet to the lowest triplet excited state in solution depend on the ligands. Larger and more flexible ligands typically promote dephasing, thereby shortening the coherence time\cite{Kim2022}. 

[Pt$_2$(P$_2$O$_5$H$_2$)$_4$]$^{4-}$ (PtPOP) is the quintessential prototype of Pt(II)-Pt(II) compound, having been extensively studied both experimentally\cite{Haldrup2019, monni2018, Biasin2018, MONNI2017, Veen2011, VanDerVeen2009, Christensen2008, Kim2002, Thiel1993, Ikeyama1988, Rice1983, Fordyce1981, Sadler1980} and theoretically\cite{Karak2021, levi2018, Zalis2015, Kong2012, Penfold2012, Stoyanov2004, Novozhilova2003, Gellene2002}. \added[id=GL1, comment=R1C2 and R2C1]{Below 320 nm, the optical absorption spectrum of PtPOP in solution exhibits a single intense band around 370 nm due to an electric dipole-allowed transition to the lowest singlet excited state\cite{gray2017, MONNI2017, Zalis2015}. Between 200 and 320 nm, only weakly-allowed UV transitions are observed\cite{gray2017, MONNI2017, Zalis2015}. Thus, ultrafast experiments on PtPOP typically involve direct excitation to the lowest singlet excited state.}
\replaced[id=GL1]{Such excitation corresponds to a p$\sigma\leftarrow\mr{d}\sigma^*$ transition, resulting}{Photoexcitation to the lowest singlet excited state, which corresponds to a p$\sigma^*\leftarrow\mr{d}\sigma^*$ transition, results} in a 0.24–0.31 Å shortening of the Pt-Pt bond in solution\cite{Haldrup2019, Biasin2018, VanDerVeen2009, Christensen2008}. This bond contraction is accompanied by remarkably long-lived coherent Pt-Pt oscillations, as observed in both the ground and excited states through ultrafast optical spectroscopy\cite{monni2018, MONNI2017, Veen2011} and X-ray scattering experiments\cite{Haldrup2019}. The decoherence time is found to be in the range of 1.5-2.5 ps\cite{Haldrup2019, monni2018, Veen2011}, for a wide range of solvents. In contrast, intersystem crossing to the lowest triplet excited state is significantly affected by the solvent, being as fast as 700-900 fs\cite{monni2018} in acetonitrile and as slow as 14-15 ps in water and 26-29 ps in ethanol\cite{Veen2011}. The faster transition to the lowest triplet state in acetonitrile has been shown to be driven by spin-vibronic coupling between the singlet excited state and an intermediate triplet state with charge transfer character, which is stabilized by this polar solvent\cite{Karak2021, monni2018}. Significantly, since decoherence is slower than intersystem crossing in acetonitrile, the vibrational coherence is preserved during the electronic transition, leading to the observation of coherent vibrations in the lowest triplet excited state\cite{Karak2021, monni2018}. 

Previous quantum mechanics/molecular mechanics (QM/MM) molecular dynamics simulations with explicit description of the solvent have provided some insights into the remarkably long photoinduced vibrational coherence of PtPOP in solution\cite{levi2018}. The structure of the complex, consisting of square planar PtP$_4$ units linked by rigid P-O-P bridging ligands, provides a highly harmonic force constant for the Pt-Pt vibrations and shields the metal atoms from collisions with solvent molecules. Therefore, pure dephasing is inefficient and decoherence is essentially driven by relaxation of the excess vibrational energy along the Pt-Pt coordinate. Despite these advancements in the understanding of the mechanism of vibrational decoherence of the electronically excited PtPOP complex in solution, several questions remain unanswered: What are the pathways of vibrational energy transfer to the solvent? Is the energy transferred directly from the Pt atoms? Does energy dissipation occur via collisions or are solute-solvent long-range electrostatic interactions also playing a role? Ultimately, what is the mechanism of vibrational energy relaxation that accounts for the exceptionally slow decoherence of the photoexcited PtPOP complex in solution?

Nonequilibrium molecular dynamics simulations can be used to analyze the pathways of energy flow from a vibrationally excited solute to a solvent based on the instantaneous atomic forces arising from the solute-solvent interactions\cite{JuradoRomero2023, Rey2012, Rey2009, Kandratsenka2009, Vikhrenko1999, Heidelbach1999, Heidelbach1998, WhitnellHynes92, Ohmine1986}. This approach is commonly referred to as energy flux, or power-work analysis, since the portion of excess excitation energy transferred to accepting degrees of freedom over a given time interval is obtained as the time integral of the power computed from the atomic forces acting on the excited or accepting degrees of freedom. The approach has been used to investigate the pathways of energy relaxation of vibrationally excited molecules in gas and condensed phase, including a water molecule excited along the bending mode in liquid water\cite{Rey2012, Rey2009}, azulene in carbon dioxide\cite{Heidelbach1999, Heidelbach1998}, nitromethane in argon and water\cite{JuradoRomero2023}, and methyl chloride in water\cite{WhitnellHynes92}. In these previous studies, the interactions between the atoms in the solute-solvent system were modelled using potential energy functions, and the vibrational modes were obtained as normal modes of vibration from the mass-weighted force constant matrix of either the vacuum equilibrium solute geometry or its instantaneous structures in solution\cite{JuradoRomero2023, Heidelbach1999, Heidelbach1998}.

In the present work, an approach is presented to analyze the flow of excess vibrational energy of a solvated molecule using instantaneous atomic forces obtained from nonequilibrium QM/MM molecular dynamics simulations, where the electronic degrees of freedom of the solute are described explicitly through electronic structure calculations and the solvent is described using potential energy functions. This makes it possible to account for the polarization of the electron density of the excited solute by the solvent. Moreover, in the approach introduced here, the vibrational modes of the excited molecule are obtained through a generalized normal mode analysis using the covariance matrix of instantaneous mass-weighted velocities\cite{Strachan2004}. This method avoids the need to compute the Hessian matrix for several atomic configurations and inherently includes finite-temperature and anharmonic effects. By projecting the atomic forces along the generalized normal modes, it is possible to decompose the vibrational energy flux into contributions by individual atoms, thus providing insights into the specific interatomic interactions that drive the transfer of energy from the excited solute to the solvent.

The energy flux analysis is applied here to nonequilibrium QM/MM adiabatic molecular dynamics simulations of PtPOP excited to the lowest singlet excited electronic state in liquid water and acetonitrile. In these simulations, the photoexcited complex is described using orbital-optimized density functional calculations\cite{Selenius2024, Schmerwitz2023, Hait2021, Levi2020}, while the solvent is modelled using potential energy functions, with the two parts of the system coupled through an embedding scheme. For both solvents, it is found that the flow of excess energy from the initially excited pinching mode to the solvent is governed by interactions between the atoms of the ligands and the surrounding solvent molecules, mainly short-range repulsive and attractive interactions represented by a Lennard-Jones potential. In water, where solvent molecules are found to transiently coordinate the Pt atoms of the complex, a significant portion of energy is transferred directly from the Pt atoms to the solvent via short-range interactions. However, a similar amount of energy is simultaneously transferred from the solvent to the Pt atoms through longer range electrostatic interactions, resulting in no net loss of energy from the Pt atoms. Since the contribution of ligand motion to the pinching mode is relatively small, the vibrational energy dissipation is inefficient. This explains the exceptionally long-lived ($\sim$1.6 ps) and solvent-independent coherent Pt-Pt oscillations of the electronically excited PtPOP complex in solution.


\section{Methodology}

\subsection{QM/MM embedding scheme}\label{sec:qmmm_model}
The molecular dynamics simulations make use of an additive QM/MM embedding scheme\cite{Dohn2020, Dohn2017, Field1990}, typically referred to as electrostatic embedding, where the total Hamiltonian of the system is given by three terms: 
\begin{equation}\label{eq:tot_ham}
{\bf \hat{H}} = {\bf \hat{H}}_{\mr{QM}} + {\bf \hat{H}}_{\mr{QM/MM}} + {\bf \hat{H}}_{\mr{MM}}
\end{equation}
The first term, ${\bf \hat{H}}_{\mr{QM}}$, is the Hamiltonian of interaction between quantum mechanically described electrons and nuclei described as classical particles. The electrons in the QM part are described using the Kohn-Sham (KS)\cite{Kohn1965, Hohenberg1964} density functional approach. 

${\bf \hat{H}}_{\mr{MM}}$ describes interparticle interactions within the MM part, as modelled using potential  energy functions. In the present case, the MM particles are represented by a fixed-value point charge force field depending only on the position of the MM particles\added[id=GL1, comment=R2C4]{, and the short-range interactions are described using a Lennard-Jones (LJ) potential}. Thus, ${\bf \hat{H}}_{\mr{MM}}$ is constant for given coordinates of atoms in the MM part and corresponds to the energy of interaction between them, ${\bf \hat{H}}_{\mr{MM}}=E_{\mr{MM}}$. 

The QM/MM interaction Hamiltonian, ${\bf \hat{H}}_{\mr{QM/MM}}$, describes electrostatic interactions (el) as well as short-range repulsion and attraction (sr) between the QM and MM particles:
\begin{equation}\label{eq:qmmm_ham}
{\bf \hat{H}}_{\mr{QM/MM}} = {\bf \hat{H}}_{\mr{el}} + {\bf \hat{H}}_{\mr{sr}} = -\sum_{m\in \mr{MM}} \dfrac{q_m}{\lvert{\bf{r}}-{\bf{R}}_m\rvert} + \sum_{\substack{m \in \mathrm{MM} \\ \alpha \in \mathrm{QM}}} \dfrac{q_m \mathcal{Z}_\alpha}{\lvert{\bf{R}}_\alpha-{\bf{R}}_m\rvert}
+ {\bf \hat{H}}_{\mr{sr}}
\end{equation}
where ${\bf{r}}$, ${\bf{R}}_\alpha$ and ${\bf{R}}_m$ are the position vectors of the electrons and nuclei in the QM part and the MM sites, respectively, and $q_m$ and $\mathcal{Z}_\alpha$ are the charges of the MM sites and QM nuclei, respectively. Here, the short-range repulsive and attractive interactions between the QM and MM atoms (or sites) are described using a Lennard-Jones \deleted[id=GL1]{(LJ)} potential:
\begin{equation}\label{eq:lj_energy}
{\bf \hat{H}}_{\mr{sr}} = E_{\mr{sr}} = \sum_{\substack{m \in \mathrm{MM} \\ \alpha \in \mathrm{QM}}} 4 \epsilon_{m\alpha} \left[ \left(\dfrac{\sigma_{m\alpha}}{\lvert{\bf{R}}_\alpha-{\bf{R}}_m\rvert}\right)^{12} - \left(\dfrac{\sigma_{m\alpha}}{\lvert{\bf{R}}_\alpha-{\bf{R}}_m\rvert}\right)^{6}\right]
\end{equation}
where $\epsilon_{m\alpha}$ is the depth of the potential energy well and $\sigma_{m\alpha}$ is the distance at which $E_{\mr{sr}}$ is zero. The electronic positions appear only in the first term of the interaction Hamiltonian, ${\bf \hat{H}}_{\mr{QM/MM}}$, on the right-hand side of eq \ref{eq:qmmm_ham}, representing the electrostatic potential of the MM particles. Therefore, the self-consistent field wave function optimization in the KS calculation is modified only by the inclusion of this additional external potential energy term, accounting for the polarization of the QM electron density by the MM charges.

The total energy of the system consists of the sum of the three terms: 
\begin{equation}\label{eq:tot_energy}
E = E_{\mr{QM}} + E_{\mr{QM/MM}} + E_{\mr{MM}}
\end{equation}
where the interaction energy, $E_{\mr{QM/MM}}$, includes the energy of electrostatic interactions, $E_{\mr{el}}$, and short-range repulsive exchange as well as attractive interactions, $E_{\mr{sr}}$, between the QM and MM particles:
\begin{equation}\label{eq:qmmm_energy}
E_{\mr{QM/MM}} = E_{\mr{el}} + E_{\mr{sr}} = -\sum_{m\in \mr{MM}} \int \dfrac{q_m n({\bf{r}})}{\lvert{\bf{r}}-{\bf{R}}_m\rvert} \mr{d} {\bf{r}}  + 
\sum_{\substack{m \in \mathrm{MM} \\ \alpha \in \mathrm{QM}}} \dfrac{q_m \mathcal{Z}_\alpha}{\lvert{\bf{R}}_\alpha-{\bf{R}}_m\rvert}
+ E_{\mr{sr}}
\end{equation}
with $n({\bf{r}})$ being the electron density of the QM part. In the present simulations, the QM part always includes only the PtPOP complex as the solute, while the MM part includes all the solvent molecules.


\subsection{Generalized normal modes}
A generalized normal mode analysis\cite{Rega2006} as presented by Strachan\cite{Strachan2004} is adopted here. According to this approach, generalized normal modes of a system of $N$ atoms at finite temperature are defined as the 3$N$ modes with velocities $\dot{Q}_i$ that satisfy the relation:
\begin{equation}\label{eq:norm_modes_corr}
  \left\langle \dot{Q}_i(t)\dot{Q}_j(t) \right\rangle \propto \delta_{ij}  \quad\quad i,j = 1,2, \dots, 3N
\end{equation}
where $\left\langle \dots \right\rangle$ indicates time averaging and $\delta_{ij}$ is the Kronecker delta. These modes can be found by diagonalizing the following covariance matrix of mass-weighted velocities:
\begin{equation}\label{eq:cov_matrix_vels}
  {\bf K} = \frac{1}{2} \left\langle \dot{{\bf q}}^\prime(t)\dot{{\bf q}}^{\prime \dagger}(t) \right\rangle
\end{equation}
where $\dot{{\bf q}}^\prime$ is a column vector of the mass-weighted velocities $\dot{q}_k^\prime = \sqrt{m_k}\dot{q}_k$, $k=1, 2, \dots, 3N$, defined in the frame of reference that translates and rotates with the molecule (body-fixed-frame velocities) as determined from the molecular dynamics simulations. Instantaneous generalized normal mode velocities are given by:
\begin{equation}\label{eq:norm_modes_vels}
\dot{{\bf Q}}(t) = {\bf L}^\dagger\dot{{\bf q}}^\prime(t)
\end{equation}
where ${\bf L}$ is the unitary matrix that diagonalizes ${\bf K}$. The generalized normal \replaced[id=GL1, comment=R2C7]{mode}{normal} velocities $\dot{Q}_i$ satisfy the condition in eq \ref{eq:norm_modes_corr}, since $\left\langle \dot{{\bf Q}}(t)\dot{{\bf Q}}^\dagger(t) \right\rangle = 2{\bf T}$,
where ${\bf T}$ is a diagonal matrix with the average kinetic energy of the generalized normal modes along the diagonal. The instantaneous kinetic energy can be expressed as a sum over the generalized normal modes: 
\begin{equation}\label{eq:tot_ekin}
T(t) = \dfrac{1}{2}\dot{{\bf Q}}^\dagger(t)\dot{{\bf Q}}(t) = \dfrac{1}{2}\sum_{i}^{3N} \dot{Q}^2_i(t) = \sum_{i}^{3N} T_i(t)
\end{equation}


\subsection{Solute-solvent energy flux}
The time variation of the kinetic energy of a generalized normal mode $i$ (power, $P_i$) is given by: 
\begin{equation}\label{eq:mode_power}
P_i(t) = \dfrac{{\mr d} T_i(t)}{{\mr d} t} = \dot{Q}_i(t) {\bf L}^{\dagger}_i \cdot \dfrac{{\mr d} \dot{{\bf q}}^\prime(t)}{{\mr d} t} = \dot{Q}_i(t) {\bf L}^{\dagger}_i \cdot {\bf F}^\prime(t)
\end{equation}
where ${\bf L}_i$ is a column vector of the transformation matrix ${\bf L}$ (see eqs \ref{eq:cov_matrix_vels} and \ref{eq:norm_modes_vels}), representing the mode $i$, and ${\bf F}^\prime(t)$ is a vector of mass-weighted atomic force components $F_k^\prime(t) = F_k(t)/\sqrt{m_k}$, $k=1, 2, \dots, 3N$, as available from the molecular dynamics simulations. A change of kinetic energy of a generalized normal mode $i$ (work, $W_i$) over a time interval $\tau$ can be computed as:
\begin{equation}\label{eq:mode_work}
W_i(\tau) \equiv \Delta T_i(\tau) = \int_0^\tau P_i(t) {\mr d} t = \int_0^\tau \dot{Q}_i(t) {\bf L}^{\dagger}_i \cdot {\bf F}^\prime(t) {\mr d} t
\end{equation}

For the molecular dynamics simulations using the QM/MM scheme presented in section \ref{sec:qmmm_model}, the instantaneous power of a generalized normal mode of the solute can be partitioned into contributions from each solute atom according to:
\begin{equation}\label{eq:mode_power_decomp}
P_i(t) = \sum_{\alpha \in \text{QM}} P^{\alpha}_i(t) = \sum_{\alpha \in \text{QM}} \dot{Q}_i(t) {\bf L}^{\alpha\dagger}_i \cdot {\bf F}^\prime_\alpha(t)
\end{equation}
where ${\bf L}^{\alpha\dagger}_i$ includes only the coefficients of the generalized normal mode transformation corresponding to solute atom $\alpha$ and ${\bf F}^\prime_\alpha(t)$ is the vector of the mass-weighted instantaneous force acting on solute atom $\alpha$. Accordingly, the kinetic energy change for mode $i$ can be partitioned into contributions from the work of each solute atom:
\begin{equation}\label{eq:mode_work_decomp}
W_i(\tau) = \sum_{\alpha \in \text{QM}} W^{\alpha}_i(\tau) = \sum_{\alpha \in \text{QM}} \int_0^\tau \dot{Q}_i(t) {\bf L}^{\alpha\dagger}_i \cdot {\bf F}^\prime_\alpha(t) {\mr d} t
\end{equation}

In the QM/MM coupling of the solute-solvent system adopted here, the force on a QM atom includes terms due to the QM interactions, ${\bf F}_\alpha^{\mr{QM}}= \added[id=GL1]{-}\partial E_{\mr{QM}}/\partial {\bf R}_{\alpha}$, the electrostatic interaction between solute and solvent, ${\bf F}_\alpha^{\mr{el}}= \added[id=GL1]{-}\partial E_{\mr{el}}/\partial {\bf R}_{\alpha}$, and other, short-range interactions between solute and solvent, ${\bf F}_\alpha^{\mr{sr}}= \added[id=GL1]{-}\partial E_{\mr{sr}}/\partial {\bf R}_{\alpha}$, respectively, where ${\bf R}_{\alpha}$ is the position vector of QM atom $\alpha$: 
\begin{equation}\label{eq:qm_forces}
{\bf F}_\alpha(t) = {\bf F}^{\mr{QM}}_\alpha(t) + {\bf F}^{\mr{el}}_\alpha(t) + {\bf F}^{\mr{sr}}_\alpha(t)
\end{equation}
Therefore, the kinetic energy change of a vibrational mode of the solute can be partitioned into contributions from the work of QM forces and the work due to the forces exerted by the solvent on the solute atoms via electrostatic and shorter range repulsive and attractive interactions:
\begin{equation}\label{eq:mode_work_decomp_forces_1}
W_i(\tau) = \sum_{\alpha \in \text{QM}} \left[ W^{\alpha,\mr{QM}}_i(\tau) + W^{\alpha,\mr{el}}_i(\tau) + W^{\alpha,\mr{sr}}_i(\tau) \right] = W^{\mr{QM}}_i(\tau) + W^{\mr{el}}_i(\tau) + W^{\mr{sr}}_i(\tau)
\end{equation}
where: 
\begin{align}\label{eq:mode_work_decomp_forces_2}
W^{\mr{QM}}_i&=\sum_{\alpha \in \text{QM}} W^{\alpha,\mr{QM}}_i  \nonumber \\ 
W^{\mr{el}}_i&=\sum_{\alpha \in \text{QM}} W^{\alpha,\mr{el}}_i \\ 
W^{\mr{sr}}_i&=\sum_{\alpha \in \text{QM}} W^{\alpha,\mr{sr}}_i \nonumber
\end{align} \\
Neglecting the polarization of the solute by the electrostatic potential of the solvent, the first term of the right-hand side of eq \ref{eq:mode_work_decomp_forces_1} includes intramolecular vibrational redistribution \added[id=GL1]{(see Appendix A)}, while the last two terms account for the transfer of energy between the vibrational mode $i$ of the solute and the solvent, i.e. the external energy flux. \added[id=GL1, comment=R2C3]{Although the external energy flux terms,  $W^{\mr{el}}_i(\tau)$ and $W^{\mr{sr}}_i(\tau)$, have been derived here using only the normal mode kinetic energy, they correspond to the fraction of total energy of a mode transferred to the solvent, as shown in Appendix A using a normal mode decomposition of the intramolecular potential energy.} 

\replaced[id=GL1]{Overall}{Therefore}, by using the instantaneous QM/MM forces acting on the solute atoms, it is possible to monitor the flux of energy between vibrational modes of the solute to the solvent, and analyze it in terms of contributions from individual atoms of the solute. In the present work, this solute-solvent energy flux approach is applied to identify pathways of vibrational energy relaxation of the PtPOP complex electronically excited in solution using nonequilibrium QM/MM molecular dynamics simulations. 


\subsection{Model parameters}
Two sets of QM/MM direct molecular dynamics simulations of the PtPOP complex in solution are performed, one with water and the other with acetonitrile as the solvent. For the simulations in water, the system includes the PtPOP complex and 2710 solvent molecules in a cubic simulation box 43.5 Å wide. The simulations in acetonitrile include the PtPOP complex and 2728 solvent molecules in a cubic box 62.2 Å wide.

In all simulations, the QM/MM partition is fixed. The QM part includes only the PtPOP complex described using KS density functional calculations, employing the BLYP functional\cite{Lee1988} and the D3 approximation of dispersion interactions\cite{Grimme2010} together with the Becke-Johnson damping function\cite{d3bj}. 
\added[id=GL1, comment= R1C1]{The BLYP functional has been found in a previous study$^{28}$ to provide a PtPOP equilibrium solution structure as well as excited state structural changes and Pt-Pt oscillation period in close agreement with experimental values, when used in QM/MM simulations. The use of a computationally efficient generalized gradient approximation functional without introducing Hartree-Fock exchange makes it possible to collect a statistically significant amount of molecular dynamics data.} The calculations use a projector augmented wave (PAW) approach\cite{paw1,paw2}, where the core electrons are frozen based on the results of reference scalar relativistic calculations of the isolated atoms, while the smooth pseudo wave functions of the valence electrons are represented with tzp and a dzp basis sets of numerical atomic orbitals\cite{Larsen2009} centered on the Pt atom and on all other atoms, respectively. The KS calculations are performed on a uniform grid within a QM box with 0.18 Å spacing between the grid points. The size of the QM box ensures at least 5.0 Å of vacuum between any atom and the box edges. The excited state calculations use a time-independent approach where the orbitals are variationally optimized in a state-specific manner\cite{Selenius2024, Schmerwitz2023, Hait2021, Levi2020}. The lowest open-shell singlet excited state\deleted[id=GL1]{, S$_1$,} is represented within the spin-restricted formalism by targeting a nonaufbau KS solution corresponding to excitation of one electron from the ground state HOMO to the LUMO\cite{Malis2020, levi2018, Himmetoglu2012}. \added[id=GL1, comment=R1C2]{As shown earlier\cite{levi2018, gray2017} and confirmed here by inspection of the molecular orbitals, the HOMO has prevalent d$\sigma^*$ antibonding character resulting from the interaction of Pt d$_{z^2}$ orbitals, while the LUMO has prevalent p$\sigma$ bonding character resulting from the interaction of Pt p$_{z}$ orbitals.} The excited state calculations are performed using a direct orbital optimization method, which employs the exponential transformation and a limited-memory symmetric rank-one quasi-Newton algorithm\cite{Levi2020, Levi2020fd}. The initial maximum overlap method\cite{Barca2018} is employed to reduce the risk of variational collapse to the ground state. 

For the MM part, water molecules are described with the TIP4P force field\cite{Jorgensen1983}, while acetonitrile molecules are described with the interaction potential of Guàrdia et al.\cite{Guardia2001} based on a rigid linear three-site molecular geometry. Since the total charge of the PtPOP complex is equal to -4, four K$^+$ counterions are included in the MM part to charge neutralize the simulation box. The coupling between the QM and MM parts makes use of the QM/MM scheme illustrated in section \ref{sec:qmmm_model}. The parameters of the LJ potential function representing the short-range QM/MM interaction (eq \ref{eq:lj_energy}) are obtained by fitting values of the binding energy calculated using the BLYP functional supplemented with the D3 approximation of dispersion interactions for a range of solute-solvent clusters as described in the \replaced[id=GL1]{Supplementary Material}{Supporting Information}. 

In the dynamics simulations, the geometry of solvent molecules and the distances between any of the hydrogen atoms and the two nearest oxygen atoms in the complex are kept rigid. These constrains are enforced using the RATTLE algorithm\cite{Andersen1983}, except for the acetonitrile molecules, where holonomic constraints for linear triatomic configurations are used\cite{Ciccotti1982}. Energy transfer from the solute to the internal vibrational modes of the solvent during the solute vibrational relaxation is typically found to be negligible\cite{Rey2015, Gertner1991}, so the use of a rigid geometry for the solvent molecules should not significantly affect the results of the excited state simulations. The four K$^+$ counterions are kept more than 16 Å away from the geometric center of the PtPOP complex by applying a repulsive harmonic potential\cite{levi2018}. 

The system is first equilibrated in the ground electronic state in the canonical ensemble at 300 K using a Langevin thermostat and a time step of 2.0 fs. In total, 100000 equilibrated ground state configurations are collected for each type of solvent, corresponding to trajectories spanning 200 ps with the thermostat applied only to the solvent molecules. For each type of solvent, 50 nonequilibrium molecular dynamics trajectories are calculated within the Born-Oppenheimer approximation, each started by exciting the complex to the lowest singlet excited state using atom coordinates and momenta obtained from configurations of the equilibrated ground state trajectories separated by at least 4 ps. 
\added[id=GL1, comment=R2C1]{Optical experiments show that the ground state is repopulated on much longer timescales, on the order of a few $\mu s$\cite{gray2017, Fordyce1981}. Moreover, previous QM/MM simulations using time-dependent DFT have shown that higher singlet excited states remain energetically well-separated from the lowest one during the dynamics\cite{monni2018}. Therefore, no internal conversion occurs within 4 ps after excitation, and the adiabatic propagation on a single excited state is justified.}
The first excited state calculation for each trajectory is initialized using the orbitals of the ground state with occupation numbers reflecting promotion of an electron from the HOMO to the LUMO, i.e. the HOMO and LUMO open-shell orbitals are assigned an occupation number of 1, while the other, closed-shell orbitals are assigned an occupation number of 2. Excited state calculations at subsequent steps in the trajectory propagation use the occupied optimized orbitals of the previous molecular dynamics step as initial guess. For the nonequilibrium molecular dynamics simulations, a time step of 2.0 fs is used, and the Langevin thermostat is applied only to solvent molecules far from the solute, while the PtPOP complex and the closest solvent molecules are not coupled to the heat bath. The thermostat is switched on gradually, increasing the friction coefficient linearly from 0 to 1 ps$^{-1}$ within a 2 Å buffer region extending radially in the simulation cell starting from a radius of 12 Å from the geometric center of the solute. Each nonequilibrium excited state trajectory spans 6 ps. \added[id=GL1]{Time-dependent radial distribution functions (RDFs) of the distance between solute atoms of a given element or a site, such as the center of mass, and solvent atoms of a given element are calculated by averaging over the nonequilibrium excited state trajectories. The time-dependent cumulative coordination number is then obtained by integrating the RDFs,  $g(d, t)$:
\begin{equation}
CN(d^{\prime}, t) = 4 \pi \dfrac{N}{V} \int_0^{d^{\prime}} g(d, t) d^2 \, \mathrm{d}d
\end{equation}
where $N$ is the number of solvent atoms and $V$ is the volume of the simulation box.}

The QM/MM direct molecular dynamics simulations are performed using the atomic simulation environment (ASE)\cite{ase-paper} and the grid-based projector augmented wave (GPAW) software\cite{Mortensen2024}, where the QM/MM embedding scheme employed here is implemented\cite{Dohn2020,Dohn2017}.

For the calculations of the generalized vibrational normal modes of PtPOP from the nonequilibrium excited state QM/MM trajectories for a given solvent, the body-fixed-frame velocities are first obtained by removing the contribution of overall rotation and translation of the complex from the Cartesian velocities, as described in ref \cite{levi2018}. Then, the covariance matrix of velocities as defined in eq \ref{eq:cov_matrix_vels} is calculated, with the average carried out over all 50 excited state trajectories and all time steps. Finally, the velocity covariance matrix is diagonalized to obtain the normal mode eigenvectors used in eqs \ref{eq:mode_power}-\ref{eq:mode_work_decomp_forces_2} to analyze the flux of vibrational energy from the photoexcited complex to the solvent. \added[id=GL1, comment=R2C2]{In the present work, the energy flux associated with the pinching mode of PtPOP is analyzed, which mostly involves a change in the distance between the Pt atoms. As shown below, this mode accepts most of the excess energy due to the photoexcitation. Since the vibrational temperature of this mode is smaller than $T=300$ K ($h\nu / k_\mathrm{B}\approx170$ K, where $\nu$ is the vibrational frequency in the ground state), the equilibrium ground state distributions of mode coordinate and kinetic energy generated through classical sampling according to the equilibration procedure described above closely approximate quantum mechanical distributions$^{73}$. Therefore, quantum effects on the initial conditions for the nonequilibrium dynamics of the pinching mode are expected to be small.}


\section{Results}

\subsection{Coherent vibrational dynamics}
As expected based on previous studies\cite{levi2018}, most of the excess vibrational energy gained by the complex by photoexcitation, $\sim$60 \% in both solvents (see the \replaced[id=GL1]{Supplementary Material}{Supporting Information}), is initially concentrated in a vibrational mode corresponding to change in the distance between the two Pt atoms (hereafter referred to as pinching normal mode). In the following we focus on the coherent dynamics and energy relaxation along this vibrational mode. 

\begin{figure}[h!]
\centering
\includegraphics[scale=0.8]{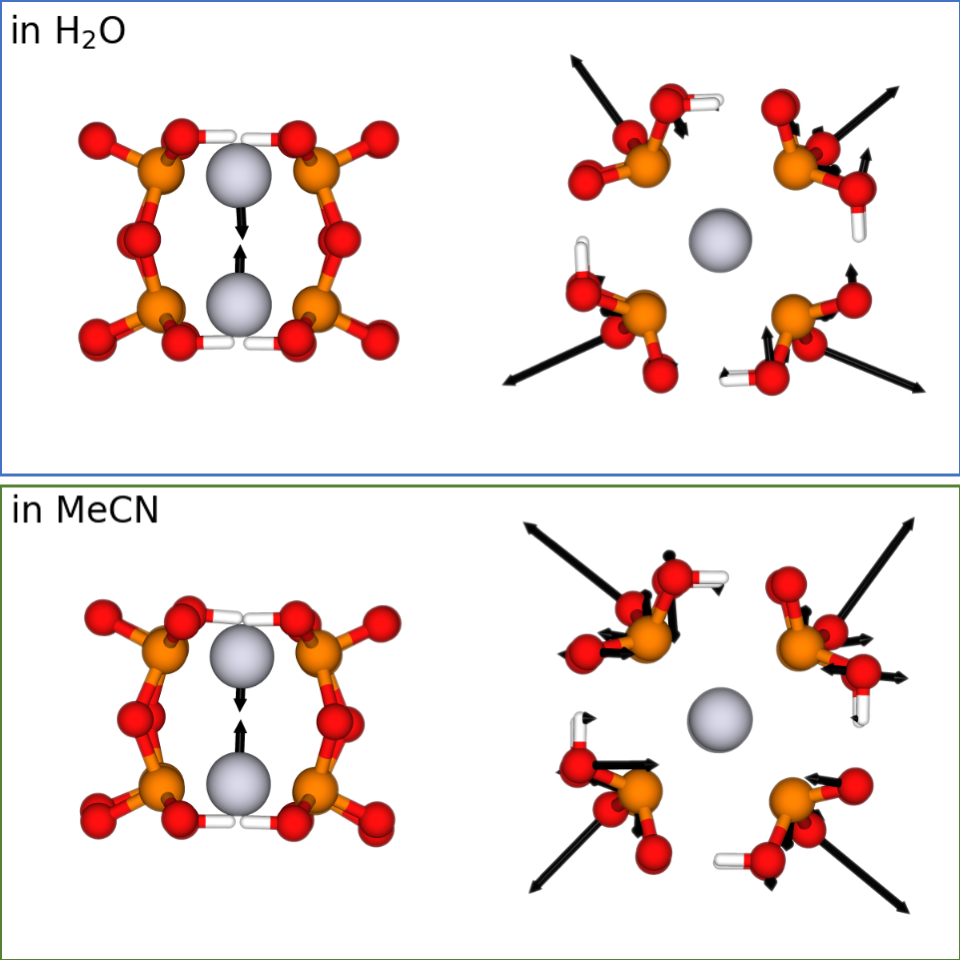}
\caption{Visualization of the PtPOP molecule with the displacement vectors of the pinching mode obtained from a generalized normal mode analysis of nonequilibrium excited state QM/MM molecular dynamics trajectories in water (top) and acetonitrile (bottom). Left: displacement vectors of the Pt atoms. Right: displacement vectors of the ligand atoms magnified by a factor of 12 for better visualization. Pt, P, O, and H atoms are grey, orange, red, and white, respectively. The pinching mode involves mostly a change in the distance between the two Pt atoms and also a small contribution of ligand bending, the latter being more pronounced in acetonitrile than in water.}
\label{fig:nmode0}
\end{figure}

Figure \ref{fig:nmode0} shows the normal mode displacement vectors obtained for both water and acetonitrile solvents. The pinching mode is dominated by oscillations in the distance between the two Pt atoms, but also involves a small contribution from the bending of the ligands: As the distance between the Pt atoms decreases (or increases) during the normal mode oscillations, the bridging O atoms are projected outwards (inwards), while the other ligand atoms are projected slightly inwards (outwards), corresponding to a decrease (increase) of the magnitude of the $\angle$P-O-P angles.  
%
%
This ligand bending contribution is more pronounced in acetonitrile than in water, pointing to greater structural rigidity in the latter solvent, likely as a result of hydrogen bonds between the water molecules in the first solvation shell and the O ligand atoms of PtPOP (see also Figure \ref{fig:hbond}). 

\begin{figure}[h!]
\includegraphics[scale=0.4]{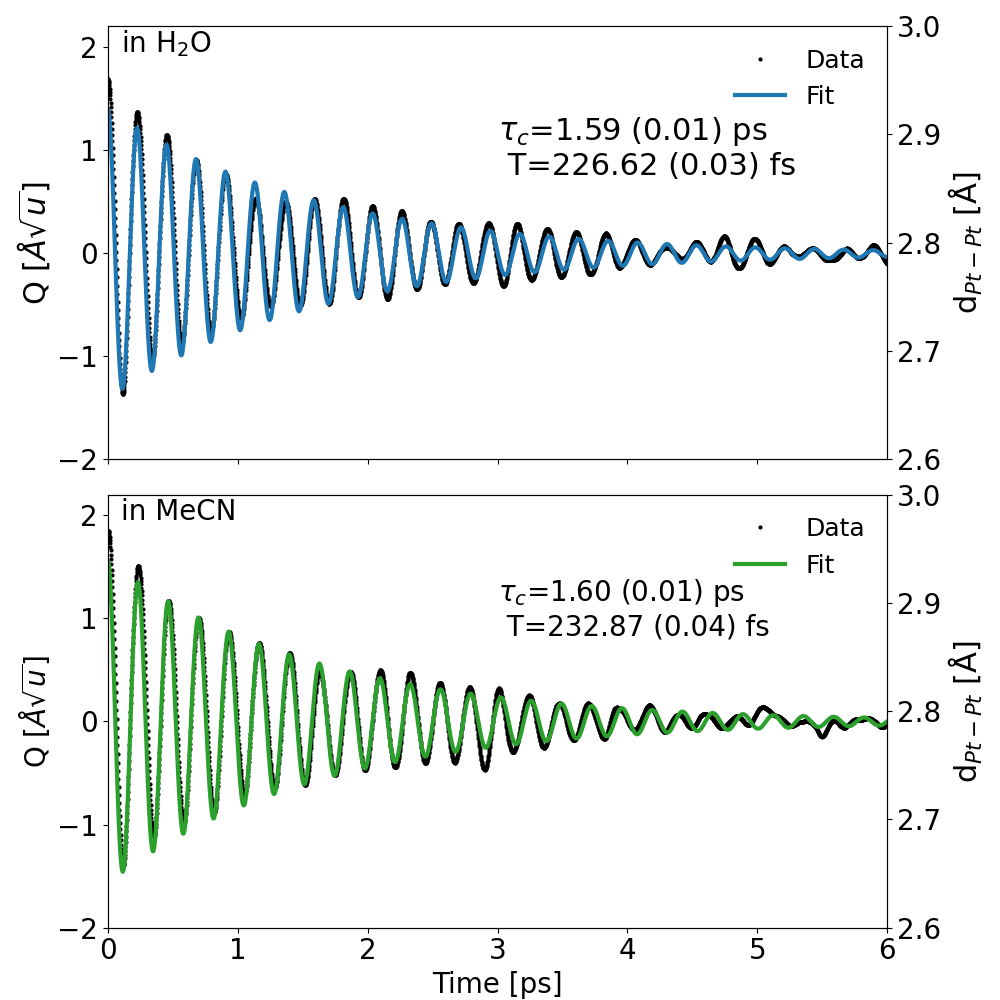}
\caption{Time evolution of the average pinching mode obtained from nonequilibrium QM/MM molecular dynamics trajectories of PtPOP excited to the lowest singlet excited state in water (top) and acetonitrile (bottom). \added[id=GL1]{The standard error of the time-dependent average Pt-Pt distance is always below 0.02 Å.} The oscillation period, $T$, and coherence decay time, $\tau_{\mr{c}}$, are obtained from fitting a periodic function with an exponentially decaying amplitude. \added[id=GL1]{The values in parentheses are the standard errors of the fits.} The decoherence time is in close agreement with the results of ultrafast transient absorption and fluorescence up-conversion experiments of PtPOP in water\cite{Veen2011} ($\tau_{\mr{c}}$ of 1.76$\pm$0.08 and 1.5$\pm$0.5 ps, respectively), where intersystem crossing occurs on longer time scale than vibrational relaxation\cite{Veen2011} ($\tau_{\mr{ISC}}>13$ ps).}
\label{fig:ddeco}
\end{figure}

Figure \ref{fig:ddeco} shows the time evolution of the pinching mode averaged over the nonequilibrium excited state trajectories for both water and acetonitrile. 
%
%
Initially, the nonequilibrium ensemble of PtPOP molecules has an average Pt-Pt bond length of \replaced[id=GL1]{$\sim$2.93 Å}{2.9 Å}, corresponding to the equilibrium bond length in the ground state. Excitation to the lowest singlet excited state induces coherent oscillations of the Pt-Pt distances in the ensemble over nearly 6 ps, after which the average Pt-Pt distance equilibrates to around \replaced[id=GL1]{2.80 Å}{2.8 Å}.  The instantaneous average normal mode coordinate is fitted with a periodic function with an exponentially decaying amplitude: 
\begin{equation}\label{eq:periodic_fit}
    f_{\mr{c}}(t) = A\textrm{e}^{-t/ \tau_{\mr{c}}} \cos{\left(\frac{2\pi}{T}t\right)}+B   
\end{equation}
where $T$ is the period of the coherent oscillations and $\tau_{\mr{c}}$ is the decoherence decay time. The fits give\deleted[id=GL1]{s} very similar oscillation periods and decoherence time for the two solvents: \replaced[id=GL1]{$T=226.62$}{$T=227$} fs and $\tau_{\mr{c}}=1.59$ ps in water, and \replaced[id=GL1]{$T=232.87$}{$T=233$} fs and $\tau_{\mr{c}}=1.60$ ps in acetonitrile. The calculated periods agree very well with the value of $\sim$224 fs obtained from previous transient absorption experiments in water\cite{Veen2011} and acetonitrile\cite{monni2018}. For water, where intersystem crossing to the lowest triplet state occurs after vibrational relaxation in the singlet excited state ($\tau_{\mr{ISC}}>$ 13 ps as determined experimentally\cite{Veen2011}), the calculated coherence decay time is very close to the values of 1.76$\pm$0.08 and 1.5$\pm$0.5 ps obtained from transient absorption and fluorescence up-conversion measurements\cite{Veen2011}, respectively. For acetonitrile, the simulations yield a slower decoherence compared to transient absorption experiments: The experimental coherence decay time for photoinduced vibrations in the lowest singlet excited state in this case is 1.1$\pm$0.1 ps\cite{monni2018}. However, for this solvent, intersystem crossing is observed to occur simultaneously to vibrational relaxation ($\tau_{\mr{ISC}}=0.7-0.9$ ps\cite{monni2018}). Therefore, the experimental value of decoherence time in acetonitrile is affected by intersystem crossing, but the present simulations do not account for this effect.

\begin{figure}[h!]
\includegraphics[scale=0.4]{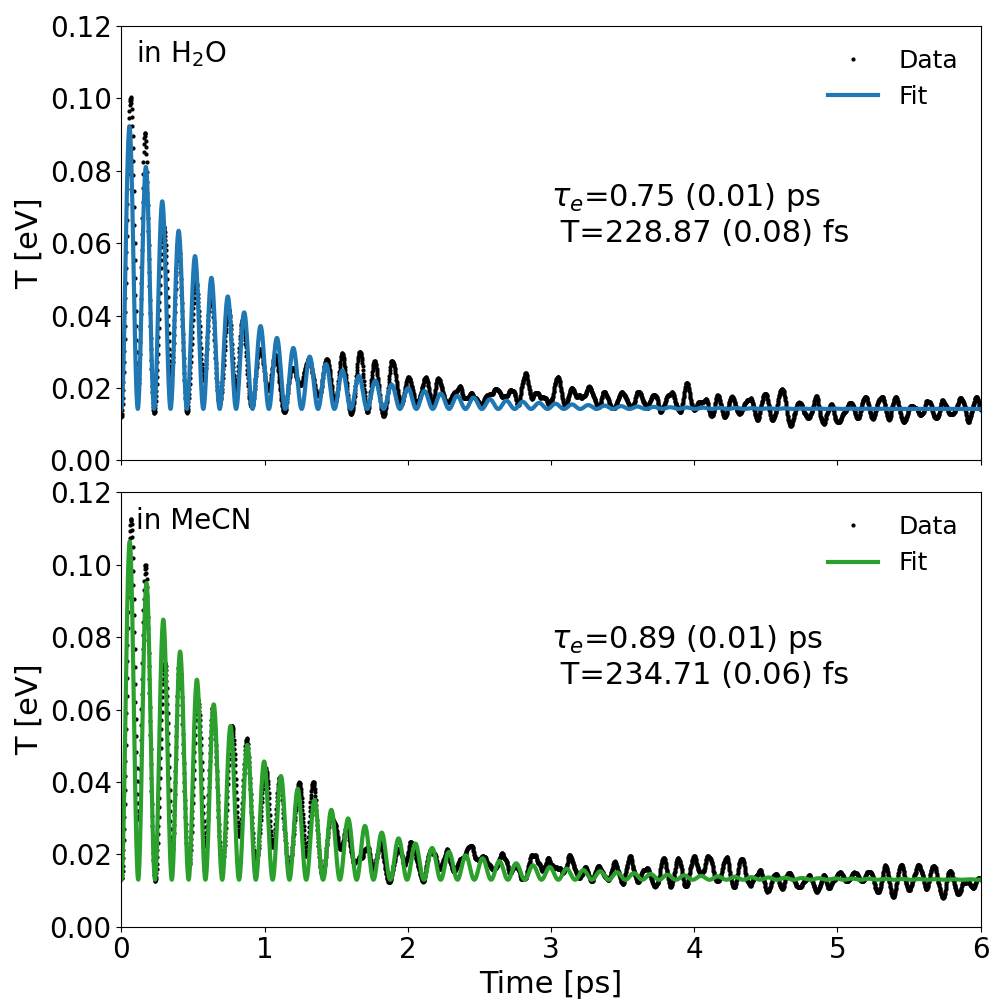}
\caption{Time evolution of the average kinetic energy in the pinching normal mode obtained from nonequilibrium QM/MM molecular dynamics trajectories of PtPOP excited to the lowest singlet excited state in water (top) and acetonitrile (bottom). \added[id=GL1]{The standard error of the time-dependent average kinetic energy is always below 0.012 eV.} The vibrational decay time, $\tau_{\mr{e}}$, is obtained from fitting a periodic function with an exponentially decaying amplitude. \added[id=GL1]{The values in parentheses are the standard errors of the fits.} The calculated $\tau_{\mr{e}}$ is almost half the coherence decay time, $\tau_{\mr{c}}$, which shows that for both water and acetonitrile vibrational coherence is lost as a result of vibrational cooling and not pure dephasing.}
\label{fig:ekin_deco}
\end{figure}
%
Figure \ref{fig:ekin_deco} shows the time evolution of the kinetic energy of the pinching mode averaged over the nonequilibrium excited state trajectories for both water and acetonitrile.
The instantaneous pinching normal mode kinetic energy is fitted with the following periodic function with an exponentially decaying amplitude:
\begin{equation}\label{eq:periodic_fit_en}
    f_{\mr{e}}(t) = C\textrm{e}^{-t/ \tau_{\mr{e}}} \cos^2 \left( \frac{2\pi}{T}t+\frac{\pi}{2} \right) +D   
\end{equation}
where $\tau_{\mr{e}}$ is the vibrational energy relaxation time, or vibrational cooling time. The values of the vibrational cooling time obtained from the \replaced[id=GL1]{fits}{fitting} are 0.75 and 0.89 ps in water and acetonitrile, respectively. Decoherence can result from both energy relaxation and pure dephasing, i.e. changes in the frequency and phase of the individual oscillators in the ensemble caused by stochastic collisions with the solvent and anharmonic effects. A decoherence time almost twice as long as the vibrational energy relaxation time indicates that the vibrational phase is to a large extent preserved during the energy relaxation and that pure dephasing effects are minimal in both solvents. As has been noted elsewhere\cite{levi2018, Veen2011}, this is likely due to the high degree of harmonicity of the Pt-Pt interaction and a cage-like ligand structure protecting the Pt-Pt oscillations from stochastic collisions with solvent molecules.


\subsection{Solvation shell structure and dynamics}\label{sec:solvation}

\begin{figure}[h!]
    
\includegraphics[width=0.95\textwidth]{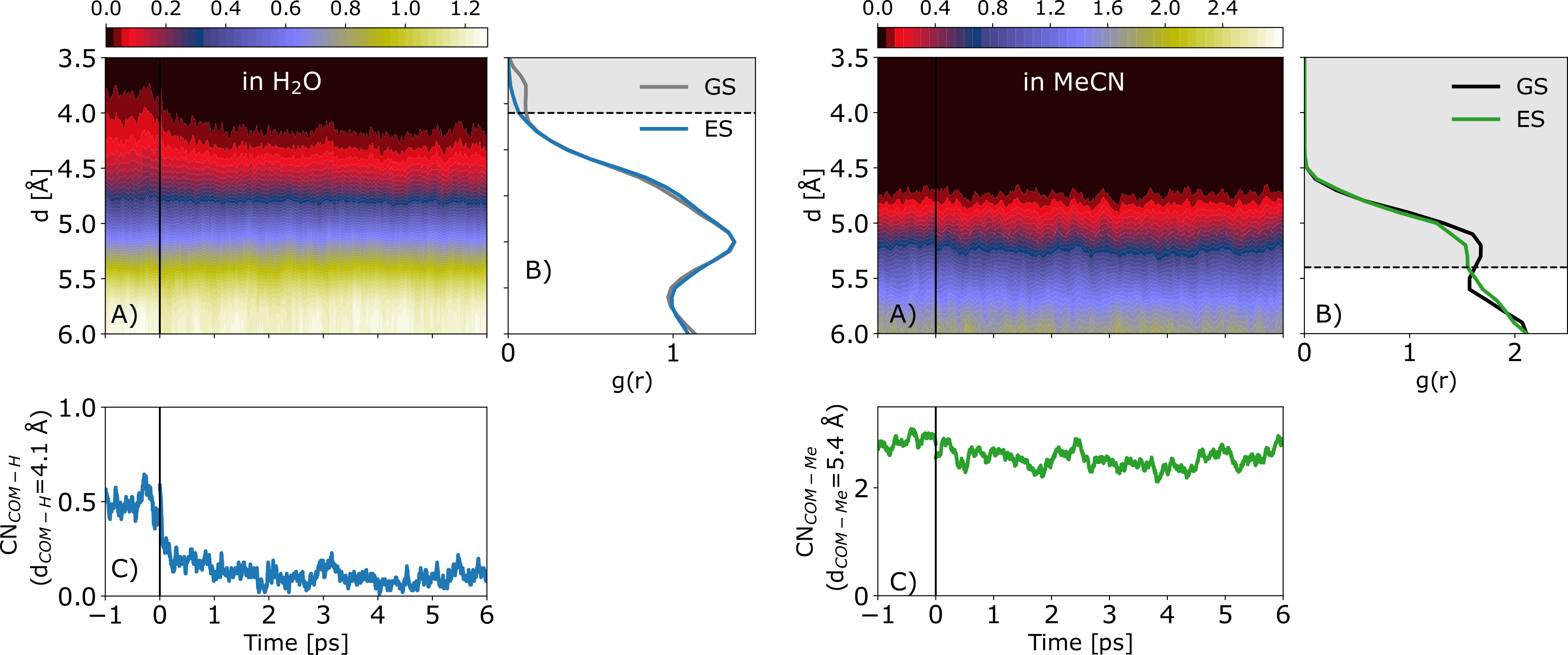}
\caption{
A) Instantaneous radial distribution functions (RDFs) of the distance between the center of mass of PtPOP and the H and methyl sites of water (left) and acetonitrile (right). The vertical black line indicates the time where the complex is excited to the lowest singlet excited state. 
B) Equilibrium ground (GS) and excited (ES) state RDFs, the latter obtained as an average over the last 2.0 ps of the nonequilibrium propagation. 
C) Instantaneous cumulative coordination number (CN) of the distance between the center of mass of PtPOP and the H and methyl sites of water and acetonitrile calculated for the distances included in the shaded regions in panels B). The solvation shell of the ground state of PtPOP in water involves a transient coordination of water molecules to the Pt atoms, which is lost after photoexcitation. The solvation shell in acetonitrile is less structured and no significant rearrangement occurs upon excitation. }
\label{fig:rdf_cum}
\end{figure}
%
Figure \ref{fig:rdf_cum} shows the \replaced[id=GL1]{RDFs}{radial distribution functions (RDFs)} of the distance between the center of mass of PtPOP and the H and methyl sites of water and acetonitrile, respectively, sampled from the thermally equilibrated ground state trajectories as well as the nonequilibrium excited state trajectories.
In the ground state of PtPOP in water, the small peak at short distance ($d<4$ Å) represents the solvation shell closest to the Pt atoms of the complex. This peak includes water molecules that transiently coordinate to the Pt atoms, with an O$-$H$\cdots$Pt preferential orientation and a Pt$-$H distance of around 2.5 Å, as visualized in Figure \ref{fig:hbond} (see also the Pt$-$H and Pt$-$O solute-solvent RDFs shown in the \replaced[id=GL1]{Supplementary Material}{Supporting Information}). 

\begin{figure}[h!]
    \includegraphics[width=0.45\textwidth]{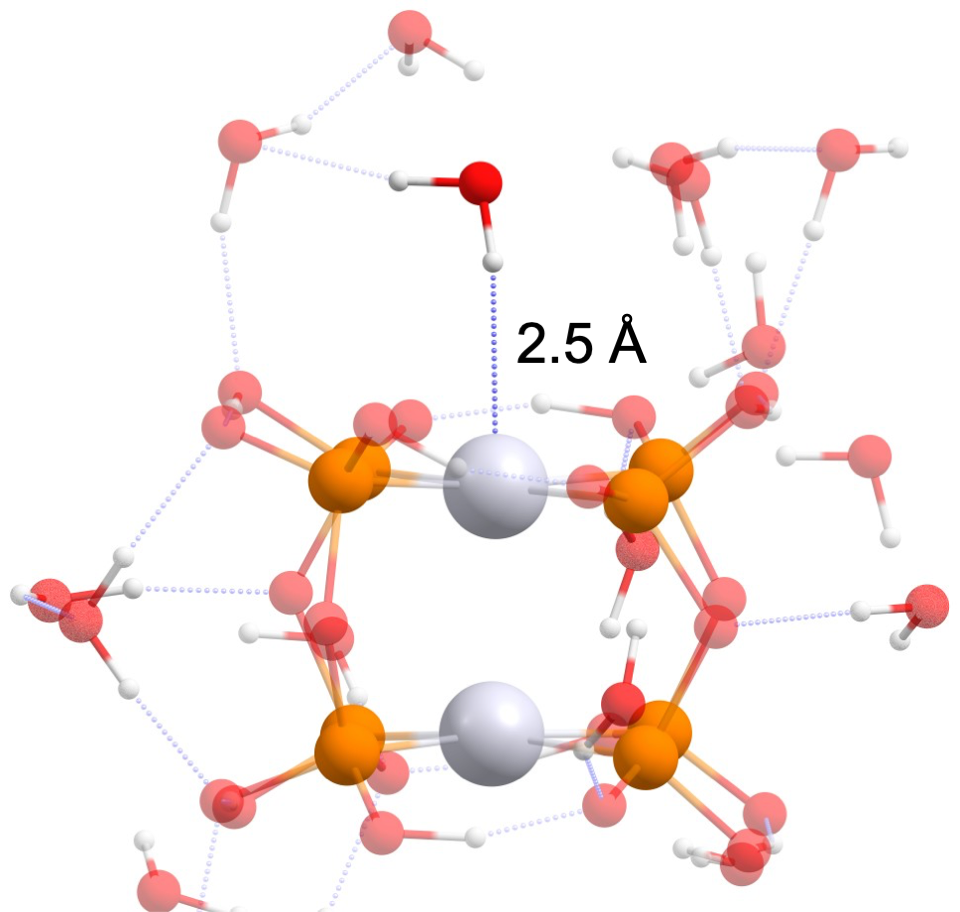}
    \caption{Visualization of a frame from an equilibrium QM/MM molecular dynamics trajectory of PtPOP in the ground state dissolved in water, highlighting the close coordination of a water molecule to a Pt atom and the hydrogen bonds involving the ligand atoms of the complex.}
    \label{fig:hbond}
\end{figure}

As shown by the instantaneous RDFs and cumulative coordination number in Figure \ref{fig:rdf_cum}, excitation to the lowest singlet excited state and contraction of the Pt-Pt distance leads to desolvation of the closely coordinating water molecules, resulting in the loss of the first coordination shell within $\sim$1 ps, corresponding to around four full periods of the Pt-Pt coherent oscillations. This behavior is similar to what has previously been observed in the excited state dynamics of the [Ir$_2$(dimen)$_4$]$^{2+}$ (dimen = diisocyano-para-menthane) complex in acetonitrile\cite{VanDriel2016, Dohn2014}, although in the latter case the initial loss of coordination of the solvent methyl groups to the Ir atoms is followed by coordination by the strongly electron donating nitrogen atoms of the solvent. The ground state solute-solvent RDFs of PtPOP in acetonitrile appear broader, indicating a lower degree of ordering in the solvation shell compared to water. No peak is observed in the Pt$-$solvent RDFs below 3 Å (see the \replaced[id=GL1]{Supplementary Material}{Supporting Information}), suggesting that there is no strong coordination between the acetonitrile molecules and the Pt atoms. Finally, for this solvent, the solute-solvent RDFs show no significant changes upon photoexcitation.


\subsection{Analysis of solute-solvent energy flux}
\begin{figure}[h!]
\includegraphics[scale=0.4]{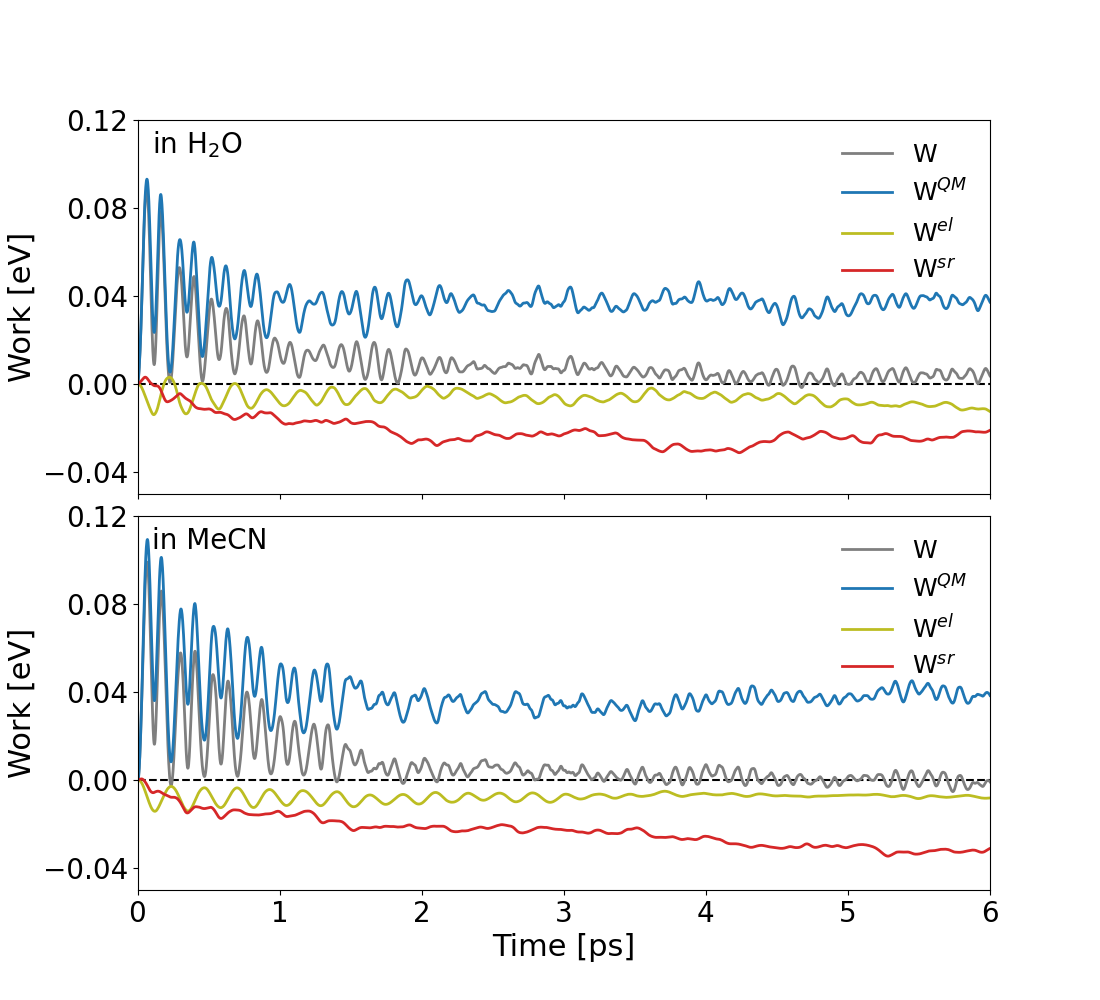}
\caption{Time evolution of the average total work associated with the pinching mode obtained from nonequilibrium QM/MM molecular dynamics trajectories of the PtPOP complex excited to the lowest singlet excited state in water (top) and acetonitrile (bottom). The contributions of the internal work due to QM forces ($W^{\mr{QM}}$), as well as the external work due to electrostatic as well as short-range repulsive and attractive Lennard-Jones (LJ) forces exerted by the solvent ($W^{\mr{el}}$ and $W^{\mr{sr}}$) are also shown. The work of electrostatic forces oscillates around a constant negative value with the same period as the oscillations of the normal mode, while the work of the short-range forces decreases monotonically. At the end of the nonequilibrium dynamics, most of the excess vibrational energy of the complex has been transferred to the solvent via the short-range LJ interactions.}
\label{fig:work_0_all}
\end{figure}

Figure \ref{fig:work_0_all} shows the instantaneous total work on the pinching mode computed as an average over the nonequilibrium excited state trajectories in water and acetonitrile, corresponding to the change $T(\tau)-T(0)$ of the instantaneous kinetic energy (see also Figure \ref{fig:ekin_deco}).
%
%
Figure \ref{fig:work_0_all} also shows how the total work is partitioned into internal work due to QM forces, $W^{\mr{QM}}$, and external work due to electrostatic as well as short-range repulsive and attractive forces exerted by the solvent, $W^{\mr{el}}$ and $W^{\mr{sr}}$, respectively, as obtained according to eqs \ref{eq:mode_work_decomp}$-$\ref{eq:mode_work_decomp_forces_2}. The work due to the QM forces remains positive at all times after excitation for both solvents. In contrast, the work due to solute-solvent electrostatic and short-range forces is negative and corresponds to the energy transferred from the normal mode to the solvent during the vibrational relaxation following electronic excitation. The work done by the short-range interactions
described using the LJ potential of eq \ref{eq:lj_energy}, decreases monotonically. Meanwhile, the work done by the solute-solvent electrostatic forces reaches a minimum after the first contraction of the Pt-Pt distances in the ensemble, and then exhibits coherent oscillations with the same period as the normal mode vibrations. Minima in the oscillations of the work of electrostatic forces correspond to minima in the oscillations of the normal mode, which are shown in Figure \ref{fig:ddeco}. This means that the electrostatic forces exerted by the solvent act to increase the distance between the PtP$_4$ groups in PtPOP, suggesting that solvation is more effective when the complex is elongated along the Pt-Pt distance. At the end of the nonequilibrium dynamics, most of the excess vibrational energy acquired in the pinching mode in the excitation, $\sim$63 \% for water and $\sim$79 \% for acetonitrile, is transferred to the solvent via the short-range LJ interactions.

\begin{figure}[h!]
\includegraphics[scale=0.4]{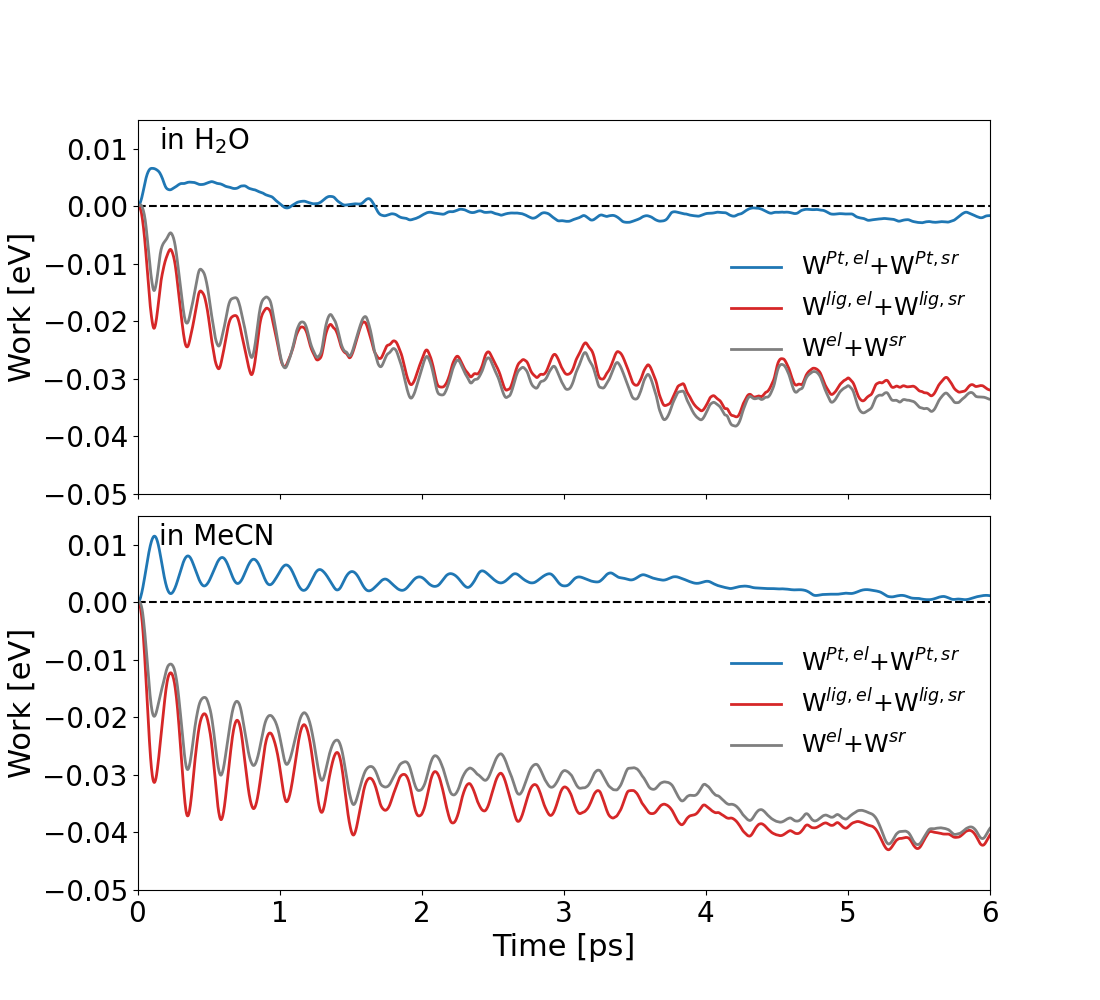}
\caption{Total instantaneous average external work on the pinching mode of PtPOP due to  electrostatic as well as short-range repulsive and attractive interactions with the solvent (the latter described by a Lennard-Jones potential), and decomposition into contributions of the Pt and ligand atoms, in water (top) and acetonitrile (bottom). Since the external work on the Pt atoms approaches zero by the end of the nonequilibrium dynamics, there is no net direct energy transfer from the Pt atoms to the solvent. Instead, the external work on ligand atoms accounts for nearly all the energy released to the solvent from the pinching mode.}
\label{fig:work_mm_pt}
\end{figure}

\begin{figure}[h!]
\includegraphics[scale=0.4]{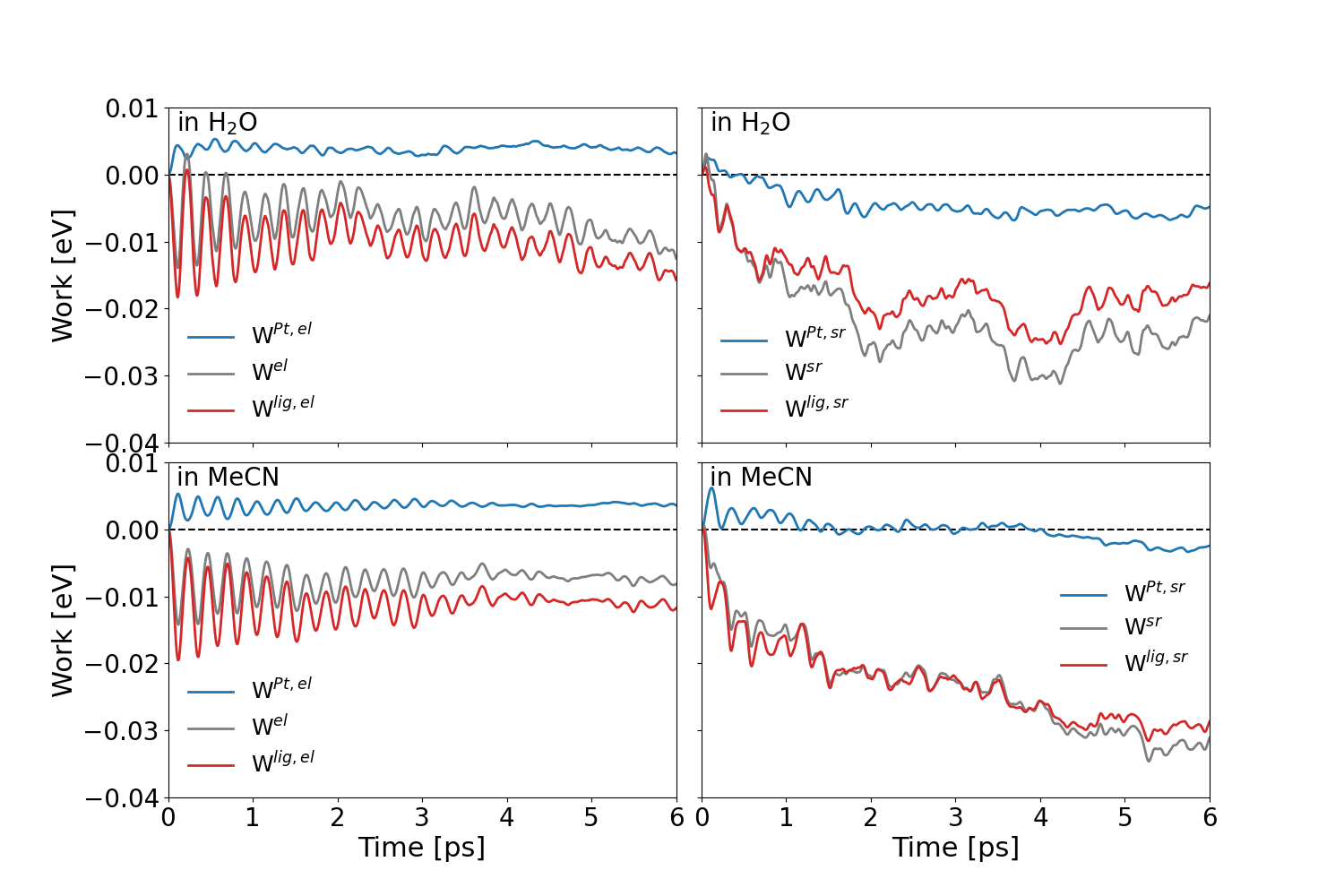}
\caption{Instantaneous average external work on the pinching mode of PtPOP due to electrostatic (left) and  short-range repulsive and attractive LJ (right) interactions with the solvent, and decomposition into contributions of the Pt and ligand atoms, in water (top) and acetonitrile (bottom). In water, a significant fraction of excess vibrational energy is released to the solvent via direct short-range interactions with the Pt atoms. However, a comparable amount of energy is received by the complex through electrostatic Pt-solvent interactions, so there is no net energy transfer via the Pt atoms (see also Figure \ref{fig:work_mm_pt}). In acetonitrile, there is no significant energy transfer to the solvent through direct Pt-solvent interactions.}
\label{fig:work_lj_pt}
\end{figure}

The contribution of the Pt and ligand atoms of the complex to the flux of energy from the pinching mode to the solvent is analyzed in Figures \ref{fig:work_mm_pt} and \ref{fig:work_lj_pt}.
%
%
Figure \ref{fig:work_mm_pt} shows that the total external work on the Pt atoms arising from interactions with the solvent is initially positive. It then decreases and approaches zero. Thus, there is no net direct energy transfer from the Pt atoms to the solvent, neither for water nor for acetonitrile. Instead, nearly all the energy transferred to the solvent is accounted for by the total external work of forces on the ligand atoms.

Figure \ref{fig:work_lj_pt} shows the time evolution of the average external work on the Pt and ligand atoms due to electrostatic interactions on the one hand and the short-range repulsive and attractive LJ interactions on the other hand.
%
%
In water, the work of short-range LJ forces on the Pt atoms accounts for a relatively large portion (approximately 23$\%$ by the end of the dynamics) of the total excess vibrational energy transferred from the excited complex to the solvent through short-range interactions. However, since the work done by the electrostatic forces on the Pt atoms is positive and of similar magnitude, these two contributions effectively cancel each other, and there is no net energy transfer directly from the Pt atoms to the solvent. This is consistent with the total instantaneous external work on the Pt atoms being close to zero, as shown in Figure \ref{fig:work_mm_pt}. Meanwhile, in acetonitrile, direct short-range LJ interactions between the Pt atoms and the solvent molecules contribute significantly less to the transfer of excess vibrational energy. Overall, both in water and acetonitrile, during the dynamics the work of short-range LJ forces on the ligands accounts for most of the total net energy released to the solvent. Electrostatic interactions with the ligands account for the rest of the energy transfer, with a bigger contribution in water compared to acetonitrile. 

\begin{figure}[h!]
\includegraphics[scale=0.4]{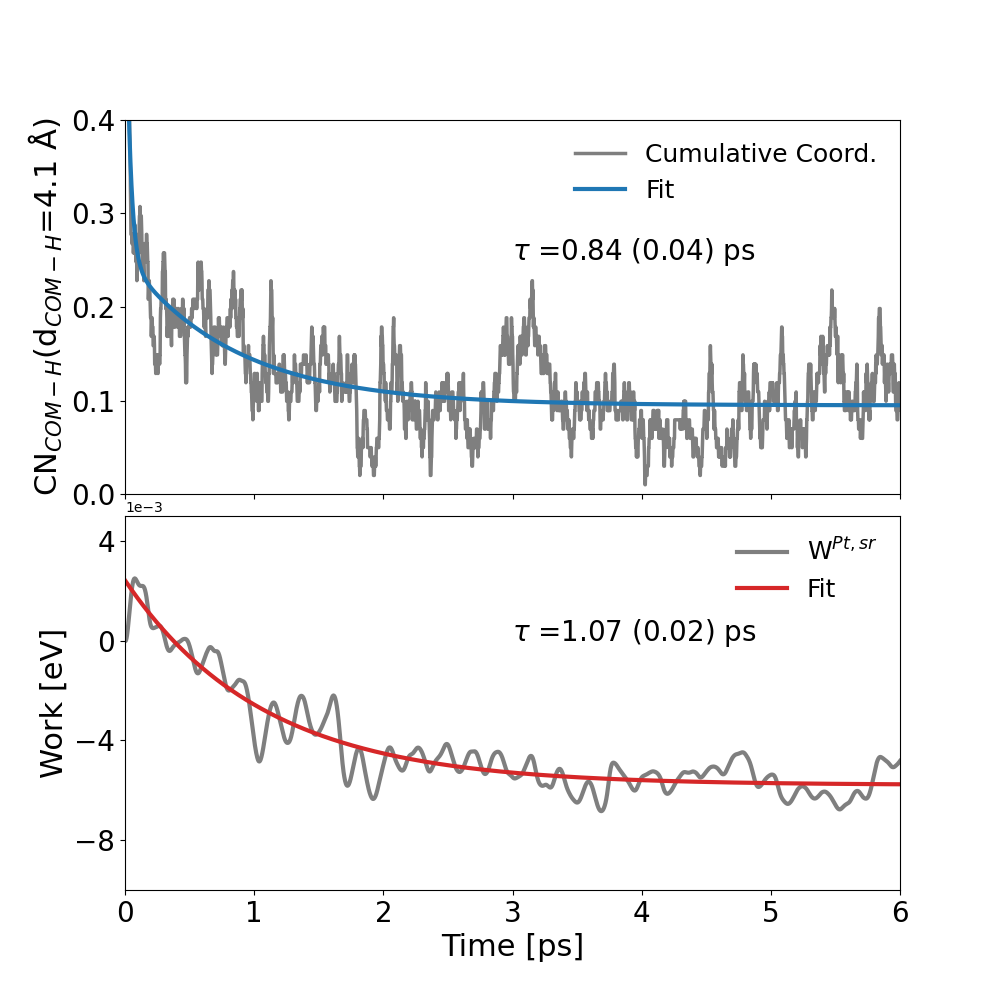}
\caption{Instantaneous cumulative coordination number (CN) calculated for distances between the center of mass of PtPOP and the H atoms of water below 4.1 Å (top) and time dependence of the component of the work on the pinching mode of PtPOP due to short-range LJ interactions between the Pt atoms and the solvent. The cumulative coordination number and the work of short-range Pt-water forces are fitted with double and single exponential  functions, respectively. \added[id=GL1]{The values in parentheses are the standard errors of the fits.} The time scales of desolvation and energy transfer via Pt-water interactions are similar, indicating that direct energy transfer from the Pt atoms to the solvent is enabled by a close coordination within the first solvation shell at the beginning of the excited state dynamics.}
\label{fig:cum_H}
\end{figure}
%
The presence of an energy dissipation channel involving direct short-range interactions between the Pt atoms and the solvent in water but not in acetonitrile can be explained by the ability of water molecules to coordinate to the metal atoms which is not the case for acetonitrile (see section \ref{sec:solvation}). The nonequilibrium ensemble created by photoexcitation initially includes configurations where water molecules strongly coordinate to the Pt atoms, exhibiting Pt$-$H distances below 3 Å, which reflects the ground state equilibrium solvation structure. As shown in section \ref{sec:solvation}, this close coordination is lost after photoexcitation. Therefore, the dissipation of excess vibrational energy from the pinching mode through direct Pt-water interactions is expected to be more efficient during the early stages of photorelaxation. This is supported by Figure \ref{fig:cum_H}, which shows that the time scale of the desolvation of water molecules in the first solvation shell is similar to the time scale of the direct energy transfer through short-range LJ Pt-water interactions.


\section{Discussion}
The energy flux approach presented above can be used to identify the pathways of energy relaxation of excited molecules in solution. The method uses the instantaneous atomic forces from nonequilibrium QM/MM molecular dynamics simulations and generalized normal modes obtained from the velocity covariances. Unlike previous methods within this category\cite{JuradoRomero2023, Rey2012, Rey2009, Kandratsenka2009, Vikhrenko1999, Heidelbach1999, Heidelbach1998, WhitnellHynes92, Ohmine1986}, where the entire system has been described using potential energy functions, the simulations presented here involve a description of the electronic structure of the solute using density functional calculations, thereby accounting for the polarization of the electron density of the solute by the environment. While the method has been applied here to analyze the flow of excess vibrational energy from the photoexcited PtPOP complex to a solvent, it can be used to analyze energy fluxes in other energy relaxation processes as well, such as the solvation relaxation following electronic excitation of a solute\cite{Rey2015}. Moreover, while the present work has focused on the contributions of the solute atoms to the energy flow, it is straightforward to extend the analysis to the identification of solvent molecules and solvent dynamical modes that accept the excess energy from the solute. 

In water, intersystem crossing from the initially excited singlet state to the lowest triplet excited state occurs more slowly than the decay time of photoinduced coherent Pt-Pt oscillations and the decoherence time obtained from the simulations (1.59 ps) agrees very well with experimental values (1.76$\pm$0.08 and 1.5$\pm$0.5 ps from transient absorption and fluorescence up-conversion measurements\cite{Veen2011}, respectively). In acetonitrile, however, intersystem crossing has been observed to occur more rapidly than vibrational decoherence, leading to the transfer of coherence from the singlet to the triplet excited state\cite{monni2018}. The coherence decay time in acetonitrile calculated here from nonequilibrium adiabatic molecular dynamics simulations (1.60 ps), which neglect intersystem crossing, is longer than the experimentally deduced singlet state decoherence time (1.1$\pm$0.1 ps), but shorter than the decoherence time observed for the lowest triplet state (2.5$\pm$0.4 ps)\cite{monni2018}. Thus, intersystem crossing appears to prevent delocalization of vibrational energy. This is further supported by the recent finding that the spin-vibronic couplings driving intersystem crossing are strongest along the pinching mode\cite{Karak2021}. The remarkable role of intersystem crossing in prolonging or even generating vibrational coherence by channelling energy along a few specific vibrational modes has been recently proposed for other Pt(II)-Pt(II) complexes as well\cite{Rafiq2023}.

The present simulations yield a decay time for the coherent Pt-Pt oscillations in the lowest singlet excited state in water that is in better agreement with experimental values compared to previous QM/MM simulations where decoherence was found to be three times faster than in the experiments\cite{levi2018}. The overestimation of the speed of decoherence in the previous simulations is likely a result of a combination of factors. Firstly, while the old simulations used the same functional (BLYP) as used here, they did not include dispersion interactions. An analysis of the potential of mean force along the Pt-Pt coordinate indicates that BLYP provides a less harmonic potential than BLYP supplemented with D3 dispersion approximation. Secondly, the previous simulations used LJ parameters from the universal force field (UFF)\cite{UFF} for the short-range repulsive and attractive QM/MM interactions instead of parameters obtained from \replaced[id=GL1]{fitting}{a fitting of} results of electronic structure calculations, as done in the present work. The solvation shell structure obtained using the UFF parameters involves a stronger transient coordination of water molecules to the Pt atoms of the complex, which is not eliminated upon excitation\cite{levi2018}. As shown here, excess Pt-Pt vibrational energy can be dissipated through the short-range interactions between the Pt atoms and surrounding water molecules. Therefore, a stronger Pt-solvent coordination can lead to a faster vibrational energy relaxation. Thirdly, the previous simulations included smearing of the occupation numbers to aid convergence of the excited state KS calculations. Such smearing has been shown to cause discontinuities along the potential energy surface when electronic states approach each other in energy\cite{Levi2020}, as can transiently occur during dynamics in solution, thereby accelerating dephasing. The present simulations use a more robust, direct optimization approach to converge the excited state orbital optimization\cite{Levi2020}, which does not necessitate of smearing of the occupation numbers, and is therefore not affected by such potential issue. Lastly, in the previous simulations, all solvent molecules in the system, including those closest to the PtPOP molecule, were coupled to the heat bath through a Langevin thermostat, which might have also contributed to accelerating the dephasing of coherent vibrations.

The analysis of the pathways of the flow of excess energy deposited by photoexcitation the pinching mode to the solvent reveals a rather complex interplay between electrostatic and short-range LJ forces. In water, where solvent molecules can transiently coordinate the Pt atoms, energy can be transferred directly from the metal atoms through short-range interactions. However, the electrostatic forces exerted by the solvent act to compress the Pt-Pt distance by performing positive work on the Pt atoms. Thus, the two types of forces have opposing effects of similar magnitude, resulting in no net transfer of energy to the solvent directly from the Pt atoms. Instead, the excess vibrational energy is predominantely dissipated to the solvent through the ligand atoms. This energy transfer occurs primarily through short-range interactions. This is unlike what has been found for the dipolar methyl chloride molecule in water from molecular dynamics simulations with potential energy functions.\cite{WhitnellHynes92} There, excess vibrational energy was found to be transferred to the solvent mainly through electrostatic interactions. 


\section{Conclusions}
Nonequilibrium adiabatic molecular dynamics simulations have been performed to investigate the vibrational decoherence and energy relaxation pathways of the bimetallic PtPOP complex following electronic excitation in liquid water on the one hand and in acetonitrile on the other hand. The simulations use a QM/MM approach and time-independent, orbital-optimized density functional calculations to describe the lowest singlet excited state of the complex. It is found that photoinduced oscillations along the Pt-Pt pinching mode have a decoherence time of approximately 1.6 ps in both solvents. In water, where intersystem crossing is slower than decoherence, this result is in excellent agreement with experimental values of coherence decay time. In acetonitrile, where intersystem crossing is faster than decoherence, the calculated decoherence time is longer than the experimental value for the lowest singlet excited state ($\sim$1.1 ps), but shorter than that for the lowest triplet excited state ($\sim$2.5 ps). This points to a remarkable role of intersystem crossing in prolonging vibrational coherence by localizing excess vibrational energy. Future work will be aimed at confirming this hypothesis through nonadiabatic QM/MM molecular dynamics simulations including singlet-triplet transitions. \added[id=GL1, comment=R2C8]{These simulations can be performed using well-established trajectory surface hopping methods in conjunction with various electronic structure methods\cite{Mai2018WCMS}, including time-independent, orbital-optimized density functional calculations of excited states\cite{Vandaele2022, Malis2022, Malis2020}, as used here within adiabatic molecular dynamics simulations}.

Key pathways of the flow of excess vibrational energy from the solute to the solvent have been identified through a power-work analysis using the QM/MM embedding atomic forces and generalized normal modes obtained from the velocity covariances. The findings show that the energy deposited by photoexcitation in the pinching mode is dissipated to the solvent through the atoms of the ligands, while there is no net energy transfer directly from the Pt atoms. Short-range repulsive and attractive solute-solvent interactions described by a Lennard-Jones potential play a bigger role than electrostatic interactions. Overall, it appears that two key factors drive the exceptionally long-lived and largely solvent-independent vibrational coherence of the PtPOP complex: (1) weak interactions between the Pt atoms and the solvent, facilitated by shielding of the metal atoms by the ligands, and (2) a rigid ligand structure, which makes dissipation of energy through the ligands inefficient. 

Future efforts will focus on identifying the solvent molecules and motions that accept the excess energy from the photoexcited PtPOP molecule. It would also be valuable to assess the impact of the solvent polarizability on the vibrational decoherence and energy relaxation pathways. This can be done by using novel QM/MM polarizable embedding models\cite{Nottoli2021, Bondanza2020, Jonsson2019, Dohn2019, Ponder2010}, which have recently been applied within molecular dynamics simulations in combination with time-independent, orbital-optimized density functional calculations of excited states.\cite{Mazzeo2023}


\section*{Supplementary Material}
\added[id=GL1]{
The Supplementary Material contains a detailed description of the procedure used to thermally equilibrate the ground state trajectories, details on the QM/MM Lennard-Jones parameter optimization, plots of solute-solvent radial distribution functions, and plots of the change in kinetic energy of the pinching mode compared to the change in kinetic energy along other modes.}

\begin{acknowledgements}
    This work was funded by the Icelandic Research Fund (grants nos. 217734 and 207014). The authors thank Sergei Egorov and Elvar Ö. Jónsson for stimulating discussions. The calculations were carried out at the Icelandic Research E-Infrastructure (IREI) facility.
\end{acknowledgements}

\section*{Data Availability Statement}
The data that support the findings of
this study are available from the
corresponding author upon reasonable
request.

\appendix

\section{\added[id=GL1]{Power from total normal mode energy}}
\added[id=GL1]{
Here, for simplicity the time dependence is omitted from all equations. Following Ref.~\onlinecite{Vikhrenko1999}, the energy of the solute is expressed using the generalized normal modes as a sum of harmonic potentials and a coupling term, $E_\mr{c}$, arising from anharmonicities and nonlinear couplings between the modes:
\begin{equation}\label{eq:intra_pot_energy}
E_{\mr{QM}} = \dfrac{1}{2} \sum_i^{3N}\omega_i^2 Q_i^2 + E_\mr{c}
\end{equation}
where $\omega_i$ is the vibrational frequency of mode $i$. Accordingly, the force along a generalized normal mode is partitioned as follows:
\begin{equation}
\tilde{F}_{i} = {\bf L}^{\dagger}_i \cdot \left( {\bf F}^\prime_\mr{QM} + {\bf F}^{\prime}_{\mr{el}} + {\bf F}^{\prime}_{\mr{es}} \right) = -\omega_i^2 Q_i + \tilde{F}_i^{\mr{c}} + {\bf L}^{\dagger}_i \cdot \left( {\bf F}^{\prime}_{\mr{el}} + {\bf F}^{\prime}_{\mr{es}} \right)
\end{equation}
where ${\bf F}^{\prime}$ are vectors of mass-weighted atomic force components, ${\bf L}_i$ is a column vector of the transformation matrix defined in eqs \ref{eq:cov_matrix_vels} and \ref{eq:norm_modes_vels}, and $\tilde{F}_i^{\mr{c}} = {\bf L}^{\dagger}_i \cdot {\bf F}^\prime_\mr{QM} + \omega_i^2 Q_i$ accounts for couplings between the modes. The time derivative of the total harmonic energy for each mode, $T_i + E_i = (\dot{Q}^2_i + \omega_i^2 Q_i^2)/2$, is then:}
\begin{align}\label{eq:tot_power}
\dfrac{{\mr d}}{{\mr d} t}\left( T_i + E_i \right) &= \dot{Q}_i \left[
- \omega_i^2 Q_i + \tilde{F}_i^{\mr{c}} +  {\bf L}^{\dagger}_i \cdot \left( {\bf F}^{\prime}_{\mr{el}} + {\bf F}^{\prime}_{\mr{es}} \right)
+ \omega_i^2 Q_i \right]\\ \nonumber
&= \dot{Q}_i \tilde{F}_i^{\mr{c}} + \dot{Q}_i {\bf L}^{\dagger}_i \cdot \left( {\bf F}^{\prime}_{\mr{el}} + {\bf F}^{\prime}_{\mr{es}} \right) \\ \nonumber 
&= P_i^{\mr{c}} + P^{\mr{el}}_i + P^{\mr{sr}}_i
\end{align}
\added[id=GL1, comment=R2C3]{
The last two terms in eq \ref{eq:tot_power}, $P^{\mr{el}}_i$ and $P^{\mr{sr}}_i$, represent the power of external electrostatic and short-range forces on a normal mode, respectively. Time integration of these terms gives the work done by the solvent on the vibrational coordinate via electrostatic and short-rage forces, $W^{\mr{el}}_i$ and $W^{\mr{sr}}_i$, which correspond to the terms in eq \ref{eq:mode_work_decomp_forces_2} derived from considering the time evolution of the kinetic energy only. $P_i^{\mr{c}}$ represents the power associated with intermode couplings. An analysis of this term would provide information on the energy flux between vibrational modes, i.e. the intramolecular vibrational redistribution. However, a separation into pairwise contributions involving energy exchange between pairs of modes is not straightforward\cite{Vikhrenko1999}. 
}

\bibliography{main}

\end{document}


\title{Supplementary Material: Decoherence and vibrational energy relaxation of the electronically excited PtPOP complex in solution} 

\author{Benedikt O. Birgisson}
\affiliation{Science Institute and Faculty of Physical Sciences, University of Iceland, Reykjavík, Iceland}
\author{Asmus Ougaard Dohn}
\affiliation{Department of Physics, Technical University of Denmark, 2800 Lyngby, Denmark}
\author{Hannes Jónsson} 
\affiliation{Science Institute and Faculty of Physical Sciences, University of Iceland, Reykjavík, Iceland}
\affiliation{Deptartment of Chemistry, Brown University, Providence, Rhode Island 02912, USA}
\author{Gianluca Levi}
\affiliation{Science Institute and Faculty of Physical Sciences, University of Iceland, Reykjavík, Iceland}
\email{giale@hi.is}

\maketitle 

\tableofcontents

\subsection{Equilibration}
A continuous ground state (GS) equilibration was performed for PtPOP solvated in water or acetonitrile with a Langevin thermostat applied to all solvent molecules and non-Coulombic interactions between solute and solvent described with Lennard-Jones (LJ) parameters from UFF\cite{UFF}. This was allowed to run for roughly 60 ps. From this continuous GS trajectory, 25 new GS trajectories were spawned at a 2 ps interval, adding dispersion corrections (D3 with BJ damping\cite{Grimme2010,d3bj}) and newly optimized LJ parameters (see section \ref{ljopt}) and equilibrated for 12 ps. Then, the thermostat was applied to all atoms and equilibrated for 8 ps in order to speed up the equilibration. Finally, the thermostat was removed from the solute and the system was equilibrated for another 8 ps to prepare the intial conditions for the subsequent non-equilibrium excited state runs. From these 25 8 ps trajectories, 50 excited state (ES) trajectories were spawned at a 4 ps interval, removing the thermostat for all atoms except for bulk solvent molecules. 6 ps of excited state dynamics was collected for each ES simulation.     

\begin{figure}[h!]
\centering
\includegraphics[scale=0.5]{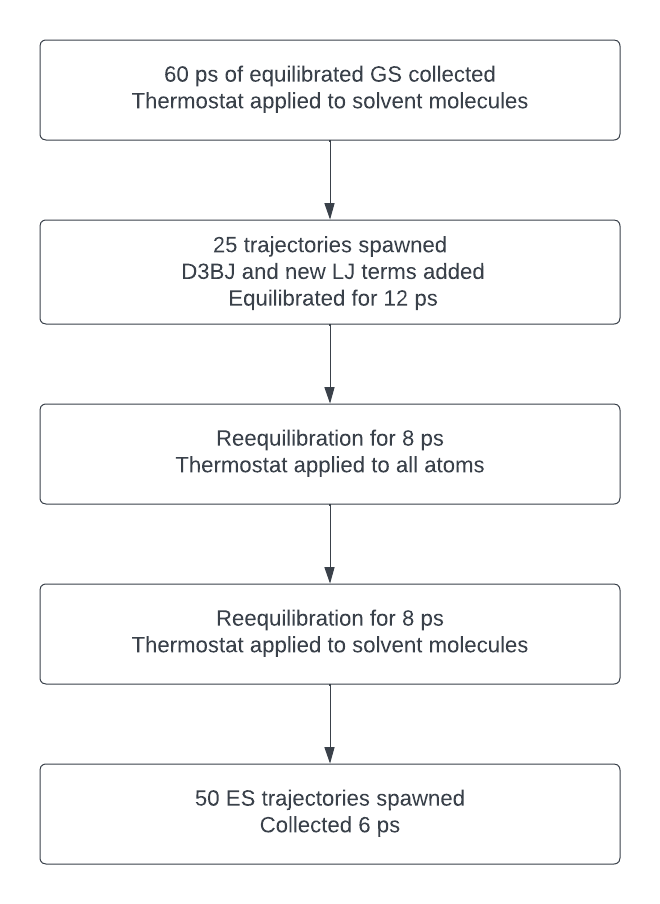}
\caption{Scheme for the equilibration procedure and subsequent excitation.}
\label{fig:equilibration}
\end{figure}

\begin{table}[H]
\centering
  \caption{Simulation Parameters}
  \label{tbl:box}
  \begin{tabular}{l| c l c }  
    &\textbf{PtPOP}  & & \textbf{PtPOP}      \\ 
  &\textbf{H$_2$O}  &   & \textbf{MeCN}      \\
 \hline
    \textbf{Solvent molecules}  & 2710 && 2728   \\
	\textbf{Counterions} & 4 K$^+$ && 4 K$^+$   \\
	\textbf{QM atoms (charge)} & 38(-4) && 38(-4) \\    	
    	\textbf{MM$_{Cell}$} & 43.486$^3$ Å$^3$ && 62.236$^3$ Å$^3$ \\
	\textbf{QM$_{Cell}$} &16.7$^3$ Å$^3$ && 17.3$^3$ Å$^3$ \\
\hline
  \end{tabular}
\end{table}

\subsection{QM/MM Lennard-Jones Parameter Optimization} \label{ljopt}
\begin{figure}[h!]
\centering
\includegraphics[scale=0.28]{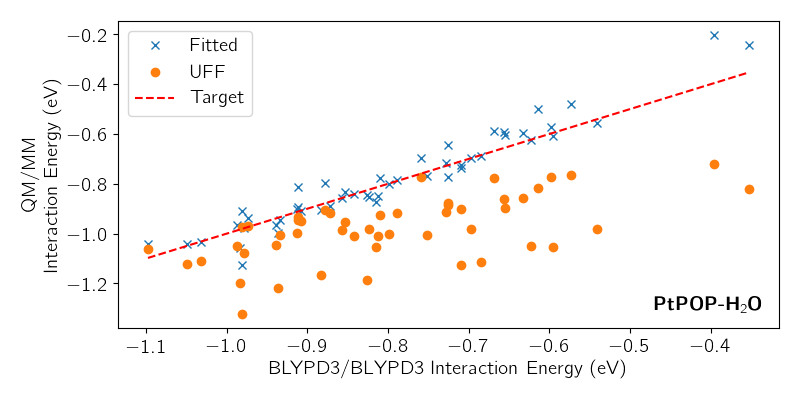}
\includegraphics[scale=0.28]{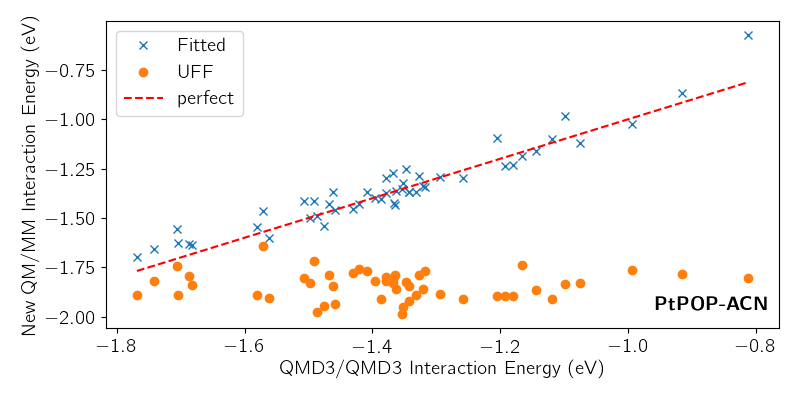}
\caption{Comparison of QM/MM interaction energies using fitted LJ parameters versus BLYP-D3/BLYP-D3 interaction energies for the selected 50 frames of the PtPOP-solvent trimer clusters for water (left) and acetonitrile (right). The orange dots represent the interaction energies obtained using the UFF LJ parameters, while the blue crosses indicate the interaction energies after optimizing their values. The red dashed line represents the ideal scenario where the QM/MM interaction energies perfectly match the BLYP-D3/BLYP-D3 interaction energies.}
\label{fig:ljfit}
\end{figure}

From the continous GS equilibration using UFF LJ parameters for the QM/MM LJ potential (see previous section), 50 frames where selected to explore as much as possible of the available configurational space, inspired by the Picky algorithm developed elsewhere\cite{Cacelli2012}. Here, we selected the 50 frames that maximize the RMSD of PtPOP-2H$_2$O (or -2MeCN) clusters by cutting out the single solvent molecule closest to each Pt atom. We then calculated the binding energies of each of these 3-molecule clusters with BLYP-D3, as well as the QM/MM binding energies, where the LJ potential was disabled. Thus, we found the set of new QM/MM LJ parameters that minimized the difference between the BLYP-D3/BLYP-D3 interaction energies and the QM/MM interaction energies, using the nonlinear least squares implemented in the scipy.optimize package\cite{2020SciPy-NMeth}. Figure \ref{fig:ljfit} shows the results of the QM/MM LJ parameter optimization.

Especially for MeCN, the UFF parameters seem to systematically overestimate the strength of the solute-solvent interactions across the range of sampled BLYPD3 interaction energies, an effect that is eliminated by the parameter optimization.

\begin{table}[H]
\centering
  \caption{PtPOP in H$_2$O Lennard-Jones Parameters}
  \label{tbl:ptp_h2o_lj}
  \begin{tabular}{l| c c  | c c }  
   & \multicolumn{2}{c|}{\textbf{Fit}} &  \multicolumn{2}{c}{\textbf{UFF}}  \\
   & \textbf{ $\epsilon$} & \textbf{$\sigma$} & \textbf{ $\epsilon$}  & \textbf{$\sigma$}  \\
  & [kcal/mol] & [Å] &  [kcal/mol] & [Å]\\
 \hline
 \hline
    \textbf{Pt}  & 0.871 & 3.621    & 0.080 & 2.454\\
	\textbf{P} & 0.305 & 3.722  &0.305 & 3.694 \\
	\textbf{O} & 0.059 & 3.461   &0.060 & 3.118 \\    	
    	\textbf{H} &  0.044 &  2.571  &0.044 &  2.571 \\
\hline
\hline
  \end{tabular}
\end{table}

\break

\begin{table}[H]
\centering
  \caption{PtPOP in MeCN Lennard-Jones Parameters}
  \label{tbl:ptp_acn_lj}
  \begin{tabular}{l| c c | c c }  
   & \multicolumn{2}{c|}{\textbf{Fit}} & \multicolumn{2}{c}{\textbf{UFF}}  \\
    & \textbf{ $\epsilon$}  & \textbf{$\sigma$} & \textbf{ $\epsilon$} & \textbf{$\sigma$}  \\
  & [kcal/mol] & [Å] & [kcal/mol] & [Å] \\
 \hline
 \hline
    \textbf{Pt}  & 0.115 & 3.430   &0.080 & 2.454\\
	\textbf{P} & 0.305 & 4.017 & 0.305 & 3.694 \\
	\textbf{O} & 0.0046 & 3.118 &0.060 & 3.118 \\    	
    	\textbf{H} &  0.008 &  2.571 &0.044 &  2.571 \\
\hline
\hline
  \end{tabular}
\end{table}

\subsection{Radial Distribution Functions}

\begin{figure}[H]
\centering
\includegraphics[scale=0.5]{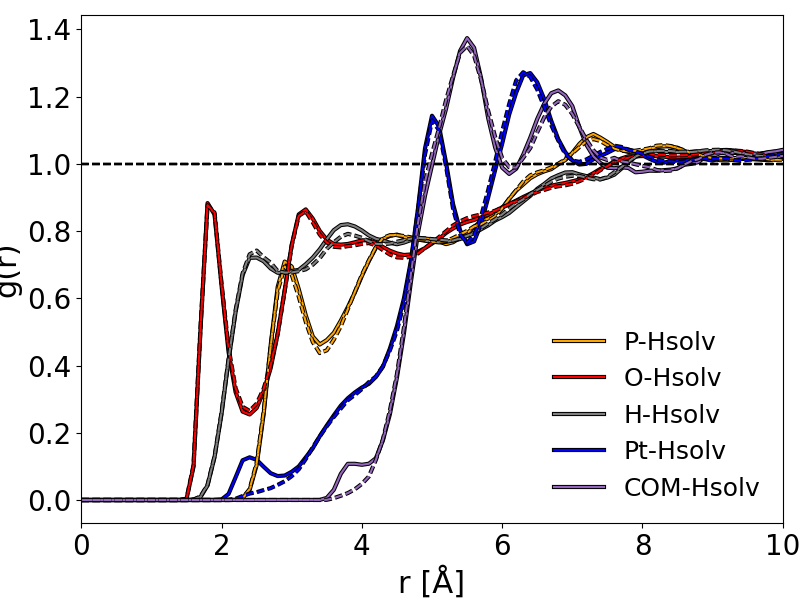}
\caption{Equilibrium RDFs of the distance between atoms of PtPOP and the hydrogen atoms of water, ES represented by dashed lines.}
\label{fig:rdf_pt_hh2o}
\end{figure}

\begin{figure}[H]
\centering
\includegraphics[scale=0.5]{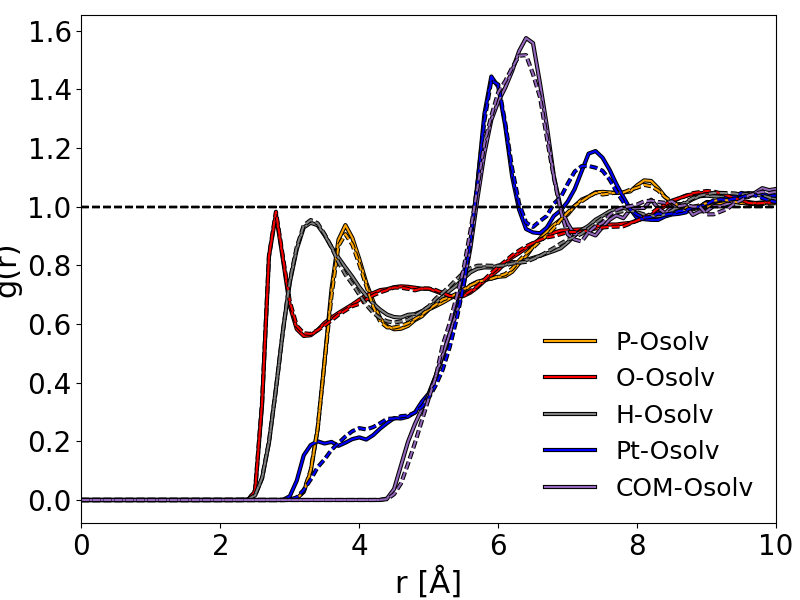}
\caption{Equilibrium RDFs of the distance between atoms of PtPOP and the oxygen atoms of water, ES represented by dashed lines.}
\label{fig:rdf_pt_oh2o}
\end{figure}

\begin{figure}[H]
\centering
\includegraphics[scale=0.5]{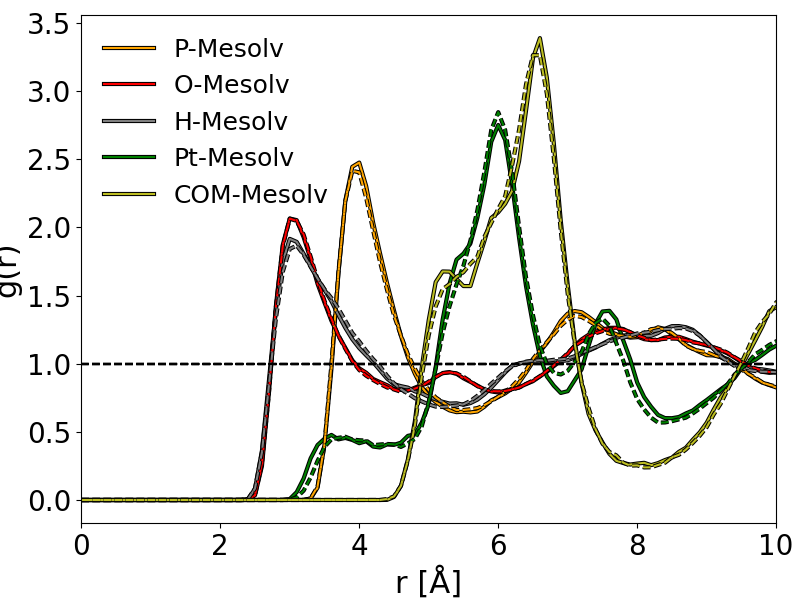}
\caption{Equilibrium RDFs of the distance between atoms of PtPOP and the methyl sites of acetonitrile, ES represented by dashed lines.}
\label{fig:rdf_pt_meacn}
\end{figure}

\begin{figure}[H]
\centering
\includegraphics[scale=0.5]{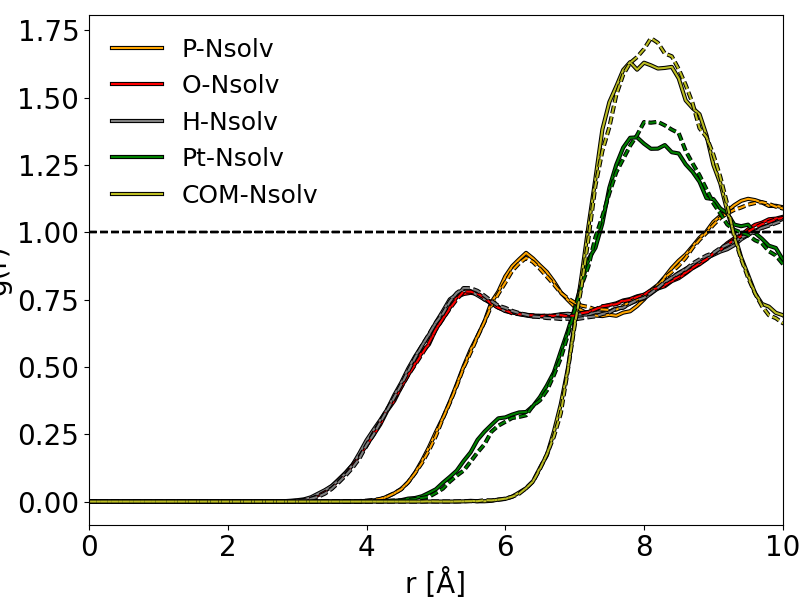}
\caption{Equilibrium RDFs of the distance between atoms of PtPOP and the nitrogen atoms of acetonitrile, ES represented by dashed lines.}
\label{fig:rdf_pt_nacn}
\end{figure}

\subsection{Time-evolution kinetic energy of normal modes}

\begin{figure}[H]
\centering
\includegraphics[scale=0.5]{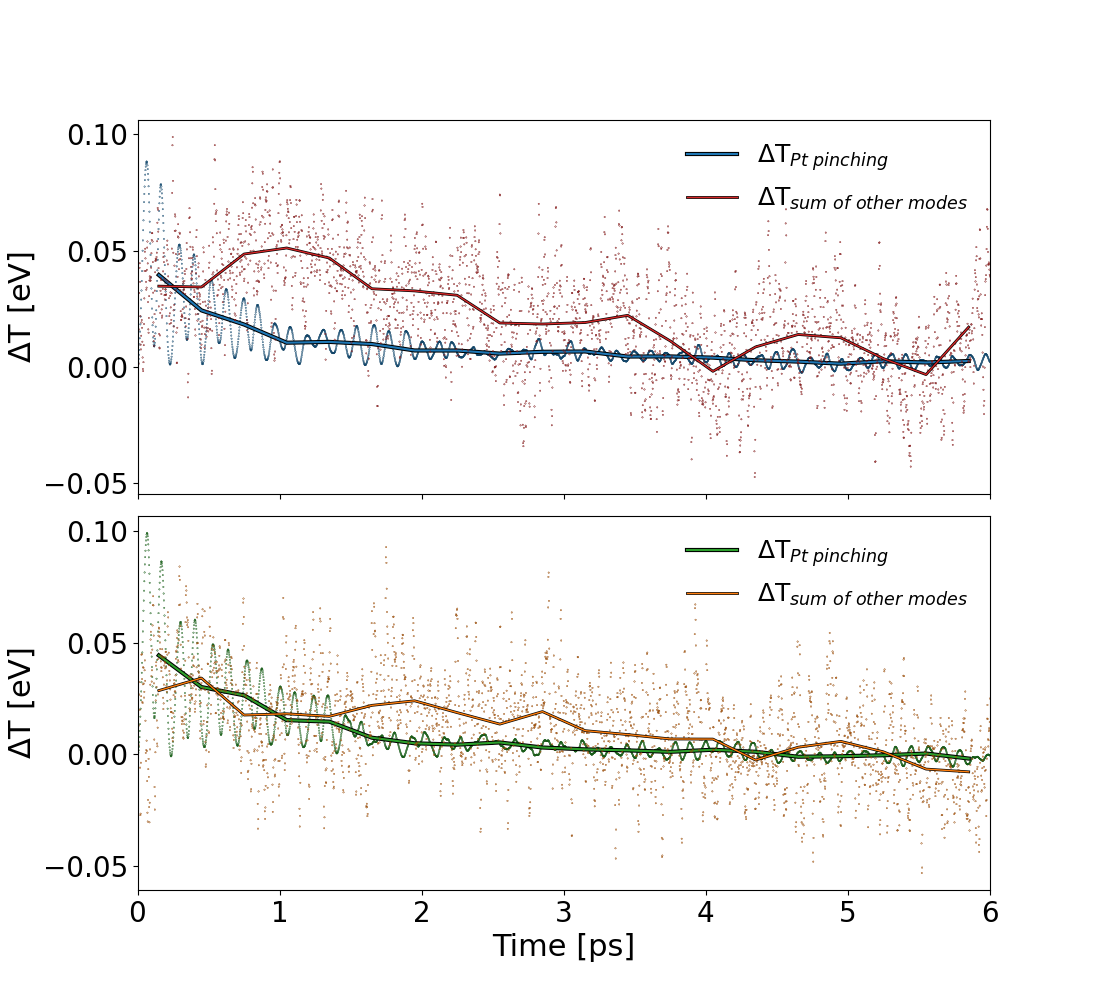}
\caption{Change in kinetic energy, $T(\tau)-T(0)$, of the pinching mode compared to all other modes after photoexcitation for water (top) and acetonitrile (bottom). The dotted lines represent the instantaneous change in kinetic energy, while the continuous lines represent an average of the instantaneous kinetic energy change over 300 fs time intervals.}
\label{fig:ediff}
\end{figure}

\break
\bibliography{si}